\documentclass[
preprint,
showpacs,preprintnumbers,
amsmath,amssymb,
aps,
prc,
longbibliography,
]{revtex4-1}

\usepackage[utf8]{inputenc}
\usepackage{graphicx}
\usepackage{dcolumn}
\usepackage{subfigure}
\usepackage{bm}
\usepackage{hyperref}
\usepackage{xcolor}

\usepackage{changes}

\makeatletter

\newcommand{\Rmnum}[1]{\expandafter\@slowromancap\romannumeral #1@}
\makeatother

\begin{document}

\title{Effects of energy levels on the double-differential cross sections of outgoing charged particles for $n + ^{19}$F reaction below 20 MeV}

\author{Hanmei Cao$^{1}$ }
\author{Fanglei Zou$^{1}$}
\email{Co-first author}
\author{Xiaojun Sun$^{1,2}$}
\email{sxj0212@gxnu.edu.cn}
\author{Jingshang Zhang$^{3}$}

\affiliation{1. College of Physics, Guangxi Normal University, Guilin 541004, People's Republic of China}
\affiliation{2. Guangxi Key Laboratory of Nuclear Physics And Nuclear Technology, Guilin, Guangxi 541004, China }         
\affiliation{3. China Nuclear Data Center, China Institute of Atomic Energy, Beijing 102413, China}
\date{\today}
\begin{abstract}
The double-differential cross sections (DDCS) for $n + ^{19}$F reaction is of critical importance for elucidating the mechanisms of nuclear reaction processes, advancing applications in nuclear engineering and technology, and supporting fundamental research in nuclear astrophysics. The quantitative description of DDCS for emission products presents a persistent theoretical challenge, primarily due to the more intricate effects of energy levels than those of 1p-shell nuclei. The pick-up mechanism of complex particles, as one of the important components of statistical theory for light nuclear reactions (STLN), is improved to describe the DDCS of outgoing charged particles, considering the effect of energy levels with energy, angular momentum and parity conservations. A comprehensive analysis of all open reaction channels is performed for $n + ^{19}$F reaction below 20 MeV. After ensuring the acquisition of high-quality DDCS of the emitted neutrons, the DDCS of outgoing charged particles (including $p, d, t, \alpha$) are self-consistently obtained. The results of this work are not only in good agreement with the recently measured experimental data at $E_n$=14.2 MeV, but also superior to the data recommended by the current major nuclear databases. Thus, LUNF code for $n + ^{19}$F reaction is developed to obtain the ENDF-6 formatted DDCS file of the nucleon and light composite charged particles. 
\end{abstract}
\pacs{25.40.-h, 24.60.Dr}
\maketitle

\section{Introduction}\label{sect1}
$^{19}$F is the only stable isotope of the element fluorine, with a natural abundance of 100\%. It demonstrates exceptional oxidizing capacity and high chemical reactivity, predominantly existing in compound forms in nature. In the realm of nuclear astrophysics, the isotope $^{19}$F takes part in nuclear reactions such as $^{19}\text{F}(n, \gamma$)$^{20}\text{F}$ and $^{19}\text{F}(n, p)^{19}\text{O}$ within high density neutron-rich environments, which include scenarios like supernova explosions and neutron star mergers. These reactions have significant implications for understanding element synthesis processes in nuclear astrophysics and elucidating cosmic nucleosynthesis mechanisms \cite{TIANTI,jose2011nuclear}.  In nuclear engineering and technology, physical quantities (such as cross section, angular distribution, double-differential cross section (DDCS), and so on) for $n + ^{19}$F reaction play a critical role in key areas including the nuclear fuel cycle, nuclear waste management, and nuclear safety systems. For example, fluoride molten salt serves as both coolants and fuel carriers in nuclear reactors, enabling the extraction, purification, recovery, and recycling of valuable nuclides such as uranium and plutonium. This facilitates efficient nuclear fuel production and reprocessing while enhancing reactor safety and economic performance \cite{yuchenggang,tuji1,tuji2,riley2019molten,jiang2022fluoride}.

The DDCS is crucial for coupling
energy and angular distributions of emitted particles, providing a concurrent description of neutron scattering and particle emission, handling numerous open reaction channels, and offering accurate distributions for particle transport, heat deposition, or radiation damage calculations \cite{ENDF-6}. Notably, the DDCS of outgoing charged particles is a fundamental value to estimate nuclear heating and material damages in fusion reactor \cite{kondo2006new}. Quantitatively describing the DDCS for neutron-induced light nucleus reactions is a challenging problem, not only due to the complexities of the reaction mechanisms, but also the scarcity of experimental DDCS of emitted charged particles. As the first nuclide in the 2s1d-shell, $^{19}\text{F}$ possesses 198 discrete energy levels \cite{tilley1995energy}, exhibiting nearly all the characteristics of 1p-shell nuclei. There are individual features of each  energy level (including energy, spin, parity, width, branching ratio, and so on), and the emission of particles between discrete energy levels is the most prominent characteristic for neutron-induced light nucleus reactions. Because of light mass for nucleon-induced light nuclear reaction, the recoil effect must be strictly taken into account.
Fortunately, the statistical theory for light nucleus reactions (STLN) was systematically described in 2015 \cite{zhang2015statistical}, based on the unified Hauser-Feshbach and exciton model. STLN has been successfully applied to calculate the double-differential cross sections of outgoing neutrons for neutron-induced reactions on $^{6}$Li \cite{jing2001model}, $^{7}$Li \cite{jing2002calculation}, $^{9}$Be \cite{junfeng2009predicted,jun2010theoretical}, $^{10}$B \cite{jing2003theoretical}, $^{11}$B \cite{jing2003theoretical1}, $^{12}$C \cite{zhang1999model,sun2008new,xiao2007analysis}, $^{13}$C \cite{Zou2025}, $^{14}$N \cite{yan2005analysis}, $^{16}$O \cite{sun2008new,jing2001theoretical,duan2005further}, and $^{19}$F \cite{jun2007theoretical}. However, the DDCS of outgoing charged particles in the aforementioned neutron-induced light nucleus reactions have hardly been published, even though the experimental measurements of the DDCS for outgoing protons, deuterons, tritons, and $\alpha$-particles for $n + ^{19}$F reaction had been performed in 2011 \cite{kondo2011measurement}. These datasets have provided a critical benchmark for theoretical calculations of charged particle emission spectra. 
 
Major international nuclear databases (including ENDF/B-VIII.1, JENDL-5, BROND-3.1, JEFF-3.3, TENDL-2023, and CENDL-3.2) have successively recommended the DDCS of outgoing products for $n + ^{19}$F reaction. In ENDF/B-VIII.1 \cite{ENDF-B-VIII.1} and BROND-3.1 \cite{BROND-3.1}, the DDCS of outgoing protons and $\alpha$-particles is evaluated using the TNG code \cite{tng}, which employs the Hauser-Feshbach multi-step statistical model. These evaluations account for reaction channels including $(n, n'), (n, p), (n, \alpha), (n, 2n), (n, np)$ and $(n, n\alpha)$. In JENDL-5 \cite{JENDL-5}, the DDCS of outgoing primary protons, deuterons, tritons, and $\alpha$ from $(n, n'), (n, p), (n, d), (n, t)$ and $(n, \alpha$) reaction channels are evaluated using the Kalbach-Mann systematics \cite{kalbach1981phenomenology,kalbach1988systematics}. For secondary emitted proton, deuteron, and $\alpha$ from $(n, np), (n, nd)$ and ($n, n\alpha$) channels, evaluations are performed using the improved Kalbach-Mann systematics \cite{kumabe1990systematics}. The file of the DDCS in JEFF-3.3 \cite{JEFF-3.3} was consistent with the ENDF/B-VII. TENDL-2023 \cite{TENDL-2023} are generally generated using the well-known Talys code \cite{Koning2023}, but for $n + ^{19}$F reaction, the data of ENDF/B-VIII.0 were directly adopted instead. The data of the DDCS derived from the major databases mentioned above, heavily relies on statistical models or systematic approaches, resulting in insufficient characterization of secondary particle emission mechanisms. On the basis of STLN, CENDL-3.2 \cite{CENDL-3.2}, including $(n, p), (n, \alpha), (n, 2n), (n, np), (n, nd), (n, nt)$ and $(n, n\alpha$) reaction channels, provided relatively accurate the DDCS of the outgoing neutrons. However, the data of the charged particles only had been preliminary estimated.

In this work,  the pick-up mechanism of the complex particles, one of the important components of STLN, is improved to describe the DDCS of the outgoing charged particles at $E_n=14.2$ MeV. In Sect. II, the theoretical framework for the DDCS of secondary particle emission between discrete energy levels is introduced for $n + ^{19}$F reaction. In Sect. III, the analysis of reaction channels below 20 MeV is performed in detail, and the DDCS of charged particles is calculated and compared with the experimental and evaluated data. A summary is succinctly given in Sect. IV.

\section{THEORETICAL MODEL}\label{sect2}
In the framework of STLN, a detailed description of the basic ideas for modeling the DDCS of outgoing nucleons and light charged particles can be found in Ref. \cite{Zou2025}. The DDCS of outgoing light charged particle $b$ with the outgoing kinetic energy $\varepsilon_b^c$ in center-of-mass system (CMS) can be expressed as \cite{Zou2025}
\begin{eqnarray}\label{eq2.1}
	\dfrac{d^2 \sigma}{d \Omega^c d \varepsilon_b^c} =  \sum_{n=3}^{n_{\max}}\dfrac{d \sigma (n)}{d \varepsilon_b^c} A(n, \varepsilon_b^c, \Omega^c)+\dfrac{1}{4\pi} \dfrac{d \sigma }{d \varepsilon_b^c}.
\end{eqnarray}
While the incident neutron energy $E_n \leq 20$ MeV, $n_{\max}=3$ can effectively describe the pre-equilibrium particle emission behavior \cite{zhang1999model}.

The energy spectrum $\dfrac{d \sigma (n)}{d \varepsilon_b^c}$ of the $n$-th exciton configuration at the pre-equilibrium stage can be expressed as
\begin{eqnarray}\label{eq2.2}
	\dfrac{d \sigma (n)}{d \varepsilon_b^c} = \sum_{j\pi}\sigma_a^{j\pi} P^{j\pi}(n)\dfrac{W_{b}^{j\pi}(n,E^*,\varepsilon_{b}^c)}{W_T^{j\pi}(n,E^*)},
\end{eqnarray}
where $\sigma_a^{j\pi}$ denotes the absorption cross section in $j\pi$ channel ($j$ and $\pi$ denote the angular momentum and parity in the final state, respectively). $P^{j\pi}(n)$ denotes the occupation probability of the $n$-th exciton configuration at the pre-equilibrium stage. $E^*$ denotes the excited energy of compound nucleus. The emission rate of the emitted particle $b$ with outgoing kinetic energy $\varepsilon_{b}^c$ of the $n$-th exciton configuration at the pre-equilibrium stage $W_{b}^{j\pi}(n,E^*,\varepsilon_{b}^c)$ can be expreesed as 
\begin{eqnarray}\label{eq2.3}
    W_{b}^{j\pi}(n,E^*,\varepsilon_{b}^c) = \sum_{k_1} W_{b,~k_1}^{j\pi}(n,E^*,\varepsilon_{b}^c).
\end{eqnarray}
Here, $W_{b,k_1}^{j\pi}(n,E^*,\varepsilon_{b}^c)$ denotes the emission rate of emitted particle $b$ with outgoing kinetic energy $\varepsilon_{b}^c$ to discrete energy levels $k_1$ of residual nucleus at the $n$-th exciton configuration, and is expressed as \begin{eqnarray}\label{eq2.4}
     W_{b,~k_1}^{j\pi}(n,E^*,\varepsilon_{b}^c)= \dfrac{1}{2\pi \hbar \omega^{j\pi}(n, E^*)} \sum_{S=|j_{k_1}-s_b|}^{j_{k_1}+s_b} \sum_{l=|j-S|}^{j+S} T_l(\varepsilon_b^c,k_1) g_l(\pi, \pi_{k_1}) F_{b\left[1, m \right]} (\varepsilon_b^c) Q_{b\left[ 1,m \right]}(n).
\end{eqnarray}
Where $\omega^{j\pi}(n,E^*)$ denotes the $n$-th exciton configuration density. $T_l(\varepsilon_b^c,k_1)$ denotes the reduced penetration factor which can be obtained by the optical model of the spherical nucleus. $j_{k_1}$ denotes the angular momentum of the residual nucleus at energy level $k_1$. $s_b$ denotes the spin of emitted particle $b$. $\pi$ and $\pi_{k_1}$ denote the parities of compound nuclei and the first residual nucleus at the $k_1$th energy level, respectively. $g_l(\pi, \pi_{k_1})$ denotes the parity conservation function. Configuration $[1, m]$  dominates the low-energy nuclear reactions \cite{Zhang2007}. $F_{b\left[1, m \right]} (\varepsilon_b^c)$ denotes the pre-formation probabilities of complex particles $b$ based on the improved Iwamoto-Harada model \cite{Duan2004}. For convenience in model calculation, the pre-formation probabilities of $[1, m]$ configuration can be calculated using the following expression
\begin{eqnarray}\label{eq2.5}
    F_{b\left[1, m \right]} (\varepsilon_b^c)= (a_1+a_2\varepsilon_f)+(b_1+b_2\varepsilon_f)\varepsilon + (c_1+c_2\varepsilon_f)\varepsilon^2,
\end{eqnarray}
where $\varepsilon = \varepsilon_b^c+B_b$, and $B_b$ denotes the binding energy of complex particle. The parameters $a_1$, $a_2$, $b_1$, $b_2$, $c_1$, $c_2$ are listed in Table 2 of Ref. \cite{Duan2004}. $Q_{b\left[ 1,m \right]}(n)$, considering the effect of the incident particle memories for neutron-induced nucleus reactions, is the combination factor of the $n$-th exciton configuration expressed as \cite{Sun2016}
\begin{gather}\label{eq2.6}
    Q_{b\left[ 1,m \right]} = \left(\dfrac{A_T}{Z_T}\right)^{z_b} \left(\dfrac{A_T}{Z_T}\right)^{n_b}  \binom{p}{\lambda}^{-1} \binom{A_T-h}{m_b}^{-1} \binom{A_b}{z_b}^{-1} \sum_{i=0}^h \binom{h}{i} \left(\dfrac{Z_T}{A_T}\right)^i \left(\dfrac{N_T}{A_T}\right)^{h-i} \\ \nonumber
	\times \sum_{j=0}^{i} \binom{i}{j} \binom{1+h-i}{\lambda-j} \binom{Z_T-i}{z_b-j} \binom{N_T-h+i}{n_b-\lambda+j}.
\end{gather}
Where $A_T$, $N_T$, and $Z_T$ denote the mass number, neutron number, and proton number of target, respectively. $n_b$ and $z_b$ denote the neutron number and proton number of complex particles, respectively. $p$ and $h$ denote the particle number and hole number at the $n$-th exciton configuration. $A_b = z_b +n_b =\lambda + m$ represents a configuration in which $\lambda$ nucleons lie above the Fermi sea and $m$ nucleons below it. The notation $\binom{.}{.}$ denotes the binomial coefficient.

The total emission rate $W_T^{j\pi}(n,E^*)$ at the pre-equilibrium stage  can be expressed as
\begin{eqnarray}\label{eq2.7}
    W_T^{j\pi}(n,E^*)=\sum_b W_{b}^{j\pi}(n,E^*,\varepsilon_{b}^c).
\end{eqnarray}

The normalized angular factor $A(n, \varepsilon_b^c, \Omega^c)$ can be expressed as
\begin{eqnarray}\label{eq2.8}
    A(n, \varepsilon_b^c, \Omega^c) = \dfrac{1}{4 \pi}\sum_{l} (2l + 1) \dfrac{G_l(\varepsilon_b^c)}{G_0(\varepsilon_b^c)} \dfrac{\tau_l(n, \varepsilon_b^c)}{\tau_0(n, \varepsilon_b^c)} P_l(\cos \theta^c),
\end{eqnarray}
where $\tau_l(n, \varepsilon_b^c)$ denotes the lifetime of the $l$-th partial wave with outgoing particle energy $\varepsilon_b^c$ emitted from the $n$-th exciton configuration, which can be derived by solving the generalized master equation. $P_l (x)$ denotes Legendre function. The geometric factor $G_l(\varepsilon_b^c)$ can be expressed as \cite{Duan2005}
\begin{eqnarray}\label{eq2.9}
	G_l({\varepsilon_b^c}) = \frac{1}{{{x_b}}}\int_{\max \left\{ {1,{x_b} - {A_b} + 1} \right\}}^{\sqrt {1 + \frac{E^*}{{{\varepsilon _F}}}} } {{x_1}d{x_1}} \int_{{x_b} - {x_1}}^{{A_b} - 1} {dy{Z_b}(y){P_l}(\cos \Theta)},
\end{eqnarray}
where $\varepsilon _F$ and $p_F$ denote Fermi energy and Fermi momentum, respectively. $x_1=p_1/p_F$, $p_1$ is momentum of the first nucleon in the outgoing composite particle $b$. And $x_b=p_b/p_F$, $p_b$ is momentum of the outgoing composite particle. $y=p_y/p_F$, and $p_y$ is total momentum of nucleons except the first nucleon in the outgoing composite particle $b$. Cosinoidal function is expressed as $ \cos \Theta = \frac{{x_b^2 + x_1^2 - {y^2}}}{{2{x_b}{x_1}}}$. ${Z_b}(y)$ is a factor related to emitted composite particle, expressed as
\begin{eqnarray}\label{eq2.10}
	{Z_b}(y) = \left\{ {\begin{array}{*{20}{l}}
			{y, ~~ ~~ ~~ ~~ ~~ ~~ ~~ ~~ ~~ ~~ ~~ ~~ ~~ ~~ ~~ ~~ ~~ ~~ ~~ ~~~~~~~~ ~~ ~~~~~~~~~b = \textmd{deuteron},}\\
			{y{{(y - 2)}^2}(y + 4), ~~ ~~ ~~ ~~ ~~ ~~ ~~ ~~ ~~~~~~~~ ~~ ~ ~~~~~~~~ ~~b = \textmd{triton}, \textmd{$^3$He},}\\
			{{{(y - 3)}^4}({y^3} + 12{y^2} + 27y - 6),~~~~~~~~~~~~~~~~~~~~b = {\alpha},}\\
			{{{(y - 4)}^6}({y^4} + 24{y^3} + 156{y^2} + 224y - 144),~~~~~b = \textmd{$^5$He}}.
	\end{array}} \right.
\end{eqnarray}
The geometric factor associated with the $n$-th exciton for nucleon is equivalent to 1, i.e., $G_l(\varepsilon_b^c)/G_0(\varepsilon_b^c)$ = 1. The formulas mentioned above are employed in this work to calculate the double-differential cross sections of outgoing neutron, proton, deuteron, triton, $^3$He, $\alpha$, and $^5$He.

The energy spectrum at the equilibrium stage $\dfrac{d \sigma}{d \varepsilon_b^c}$ can be expressed as
\begin{eqnarray}\label{eq2.11}
	\dfrac{d \sigma}{d \varepsilon_b^c} = \sum_{j\pi}\sigma_a^{j\pi} Q^{j\pi}\dfrac{W_{b}^{j\pi}(E^*,\varepsilon_{b}^c)}{W_T^{j\pi}(E^*)},
\end{eqnarray}
where $Q^{j\pi}=1-\sum_{n=3}^{n_{\max}}P^{j\pi}(n)$ denotes the occupation probability of equilibrium stage. $W_{b}^{j\pi}(E^*,\varepsilon_{b}^c)$ and $W_{T}^{j\pi}(E^*)$ denote the emission rate of particle $b$ with outgoing energy $\varepsilon_b^c$, and the total emission rate at the equilibrium stage, respectively. Their expressions are given by
\begin{eqnarray}\label{eq2.12}
    W_{T}^{j\pi}(E^*) = \sum_b W_{b}^{j\pi}(E^*,\varepsilon_{b}^c),
\end{eqnarray}
\begin{eqnarray}\label{eq2.13}
    W_{b}^{j\pi}(E^*,\varepsilon_{b}^c) = \sum_{k_1} W_{b,k_1}^{j\pi}(E^*,\varepsilon_{b}^c),
\end{eqnarray}
\begin{eqnarray}\label{eq2.14}
    W_{b,k_1}^{j\pi}(E^*,\varepsilon_{b}^c) = \dfrac{1}{2\pi \hbar \rho^{j\pi}(E^*)} \sum_{J=|j_-I_{M_1}|}^{j+I_{M_1}} \sum_{l=|J-s_b|}^{J+s_{b}} T_{Jl}(\varepsilon_b^c,k_1) g_l(\pi, \pi_{k_1}).
\end{eqnarray}
Where $\rho^{j\pi}(E^*)$ denotes the energy level density. $I_{M_1}$ denotes the spins of the first residual nucleus, and $T_{Jl}(\varepsilon_b^c,k_1)$ denotes the penetration factor.

\section{RESULTS AND ANALYSIS}\label{sect3}
\subsection{ANALYSIS OF REACTION CHANNELS}
The opening of reaction channels, their corresponding reaction $Q$ values, and the threshold energy $E_{th}$ significantly influence the DDCS of the emitted particles. Considering the emission processes from a compound nucleus to the discrete levels of the first residual nuclei, and subsequently from these levels to the discrete levels of the secondary residual nuclei, with conservations of angular momentum and parity for  $n + ^{19}$F reaction. The following reaction channels are accessible at incident neutron energies $E_n \leq 20$ MeV
\begin{equation}
    n+^{19}\mathrm{F} \to ^{20}\mathrm{F}^*\begin{cases}
     &\gamma+^{20}\mathrm{F}~~~~~~~~~~~~~~~~~~~~~~~~~~~~~~~~~~~~~~~~~~(n, \gamma),\\
  & n+^{19}\mathrm{F} \begin{cases}
  &^{19}\mathrm{F}~~~~~~~~~~~~~~~~~~~~~~~~~~~~~~~~(n, n'),\\
  & n+^{18}\mathrm{F} (k_1\ge104 )~~~~~~~~~~~~(n, 2n),\\
  & p+^{18}\mathrm{O}(k_1\ge54 )~~~~~~~~~~~~~~ (n, np),\\
  & d+^{17}\mathrm{O}(k_1\ge151 )~~~~~~~~~~~~(n, nd),\\
  & t+^{16}\mathrm{O}(k_1\ge 130)~~~~~~~~~~~~~(n, nt),\\
  & \alpha +^{15}\mathrm{N}(k_1\ge 9)~~~~~~~~~~~~~~~~(n, n\alpha),
\end{cases}  \\
& p+^{19}\mathrm{O} \begin{cases}
  & ^{19}\mathrm{O} ~~~~~~~~~~~~~~~~~~~~~~~~~~~~~~~~~ (n, p), \\
  & n+^{18}\mathrm{O}  (k_1\ge9 )~~~~~~~~~~~~~~~~~(n, pn),
 \end{cases}\\
  & d+^{18}\mathrm{O}\begin{cases}
  & ^{18}\mathrm{O} ~~~~~~~~~~~~~~~~~~~~~~~~~~~~~~~~~(n, d), \\
  & n+^{17}\mathrm{O}(k_1\ge21 )~~~~~~~~~~~~~~~(n, dn),\\
  & \alpha +^{14}\mathrm{C}(k_1\ge 12)~~~~~~~~~~~~~~~(n, d\alpha ),
 \end{cases}\\ 
  & t+^{17}\mathrm{O}\begin{cases}
  & ^{17}\mathrm{O} ~~~~~~~~~~~~~~~~~~~~~~~~~~~~~~~~~~~(n, t), \\
  & n+^{16}\mathrm{O}(k_1\ge4 )~~~~~~~~~~~~~~~~~~~(n, tn), \end{cases}\\
  & \alpha +^{16}\mathrm{N} \begin{cases}
  & ^{16}\mathrm{N}  ~~~~~~~~~~~~~~~~~~~~~~~~~~~~~~~~~~(n, \alpha ),\\
  & n +^{15}\mathrm{N}(k_1\ge 4)~~~~~~~~~~~~~~~~~~(n, \alpha n),\\
  & d +^{14}\mathrm{C}(k_1\ge50 )~~~~~~~~~~~~~~~~~(n, \alpha d),
 \end{cases}\\ 
&^5\mathrm{He} +^{15}\mathrm{N} \to n+\alpha +^{15}\mathrm{N}.
\end{cases}
\end{equation}
Here, $k_1$ denotes the energy level of the first residual nucleus and the secondary residual nucleus. The structural information of the target, compound nucleus, and first and secondary residual nuclei is taken from Refs. \cite{tilley1993energy,tilley1995energy,tilley1998energy,AJZENBERGSELOVE1991}.

After taking into account the contributions of the aforementioned reaction channels, the calculated DDCS of emitted neutrons is not only in good agreement with experimental data, but also is nearly consistent with the results of our previous work \cite{jun2007theoretical}. In this work, only the values greater than $10^{-3}$ mb/sr/MeV are presented. 

\subsection{DDCS of outgoing proton}
\begin{figure}[!htp]
    \centering
    \includegraphics[width=0.6\hsize]{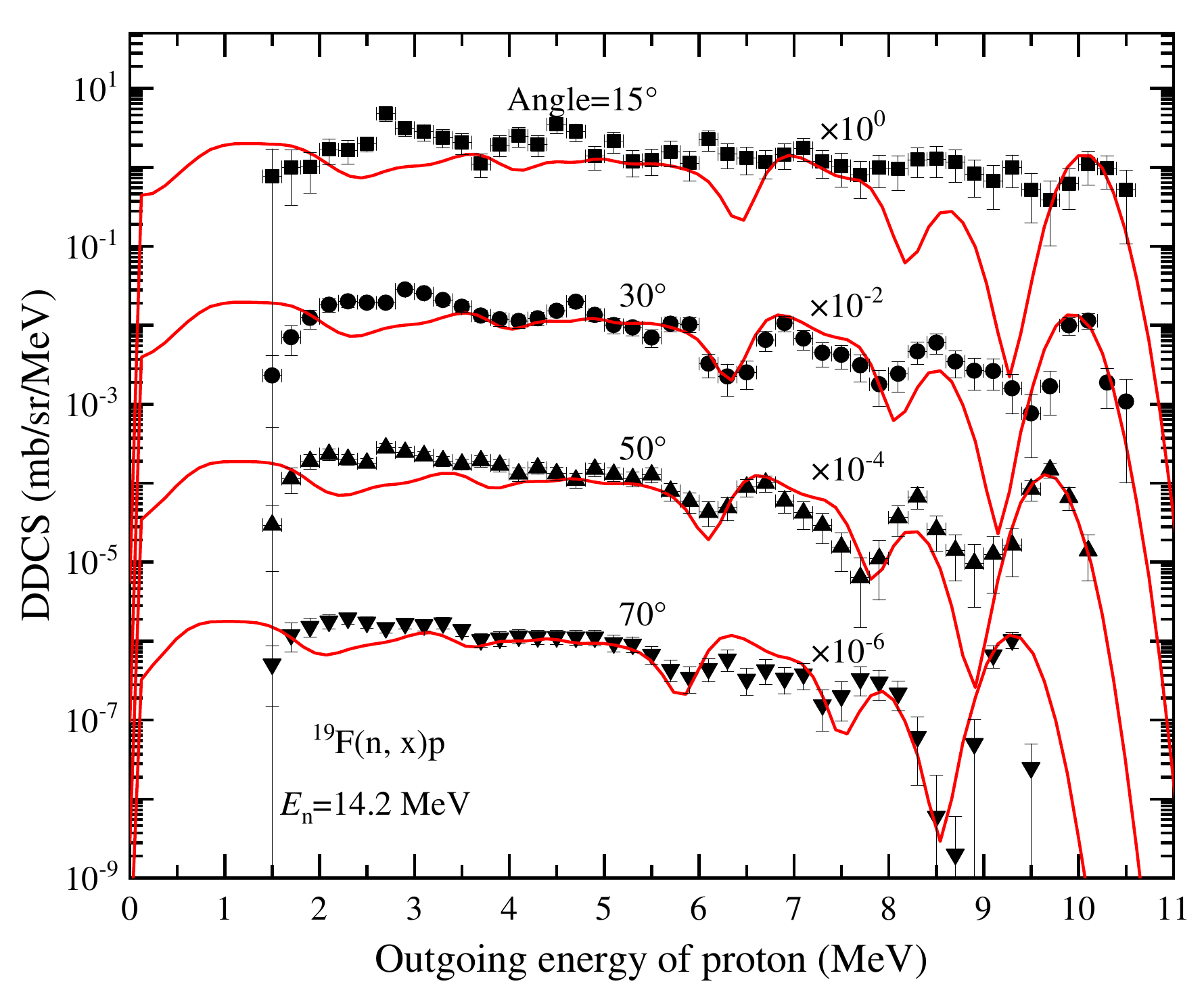}
    \caption{(Color online) Total DDCS of the outgoing proton for $\text{n} + ^{19}\mathrm{F}$ reaction with outgoing angle $15^{\circ}$, $30^{\circ}$, $50^{\circ}$, and $70^{\circ}$ at $E_n$ = 14.2 MeV in laboratory system (LS). The black points denote the experimental data taken from Ref. \cite{kondo2011measurement}, and the red solid lines denote the results of this work.}
    \label{Fig1}
\end{figure}

\begin{figure}[!htp]
    \centering
    \includegraphics[width=0.6\hsize]{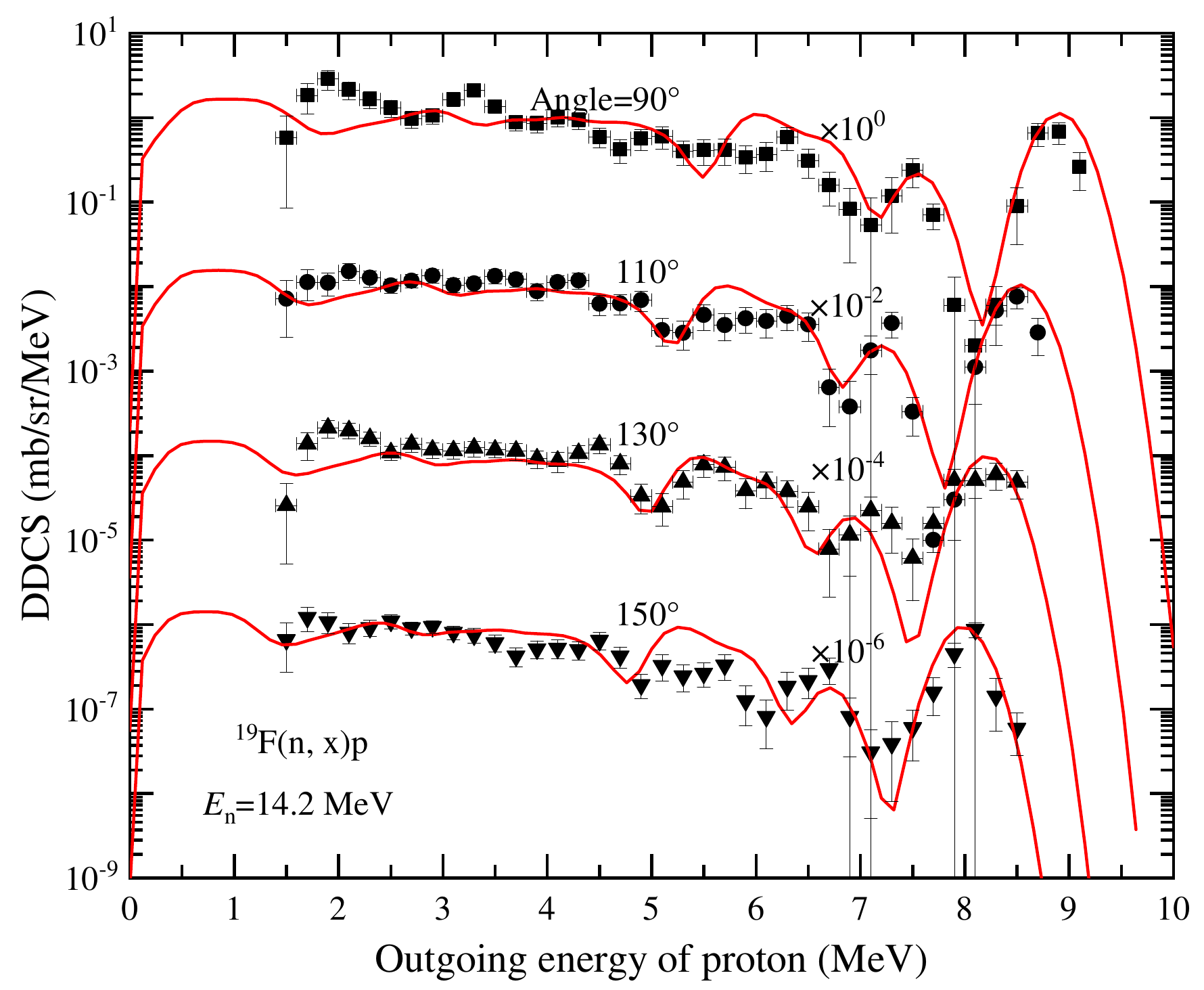}
    \caption{(Color online) Same as Fig. \ref{Fig1}, but for outgoing angles $90^{\circ}$, $110^{\circ}$, $130^{\circ}$, and $150^{\circ}$.}
    \label{Fig2}
\end{figure}

Comparisons are performed between the calculations and the measurements of the total DDCS of the outgoing proton at an incident energy $E_n = 14.2$ MeV. The results are shown in Figs. \ref{Fig1} and \ref{Fig2} for outgoing angles of $15^{\circ}$, $30^{\circ}$, $50^{\circ}$, $70^{\circ}$, $90^{\circ}$, $110^{\circ}$, $130^{\circ}$, and $150^{\circ}$, respectively. The black points represent the experimental data taken from Ref. \cite{kondo2011measurement}, and the red solid lines represent the results of this work. Data are successively shifted downward by factors of $10^{0}$, $10^{-2}$, $10^{-4}$, and $10^{-6}$ for clarity. One can see that the calculated results agree reasonably well with the experimental data. At 15$^\circ$ angle in 8–9 MeV outgoing energy region, the agreement between theoretical calculations and experimental data is slightly inferior. This is primarily due to the fact that the distance between the detector's position and the sample at 15$^\circ$ is greater than that at other angles, as shown in Table 5 of Ref. \cite{kondo2011measurement}. This case is same for the outgoing deuteron and triton.

\begin{figure}[!htp]
    \centering
    \subfigure{
        \label{Fig3a}
        \includegraphics[width=0.45\hsize]{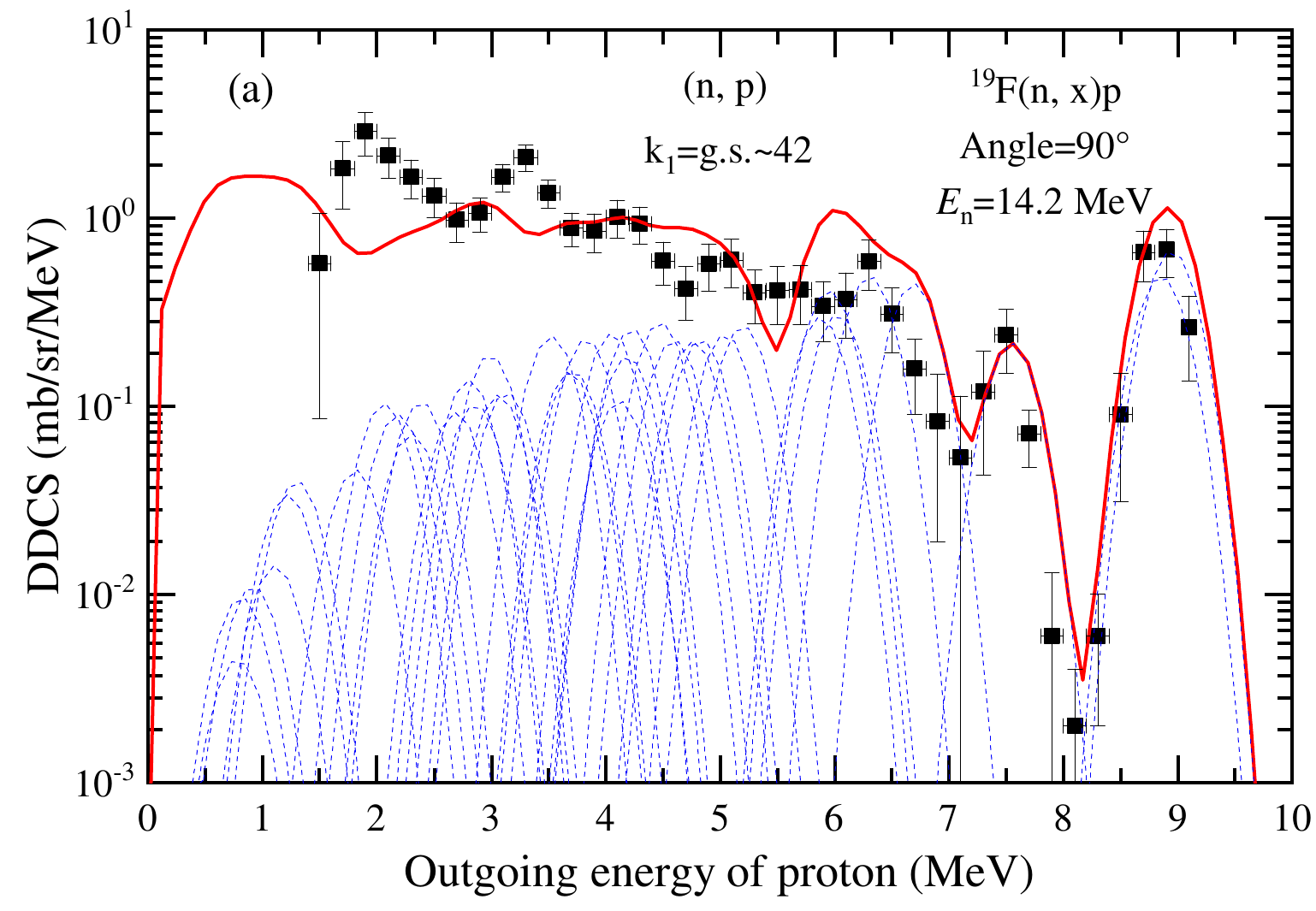}
	}
    \subfigure{
		\label{Fig3b}
		\includegraphics[width=0.45\hsize]{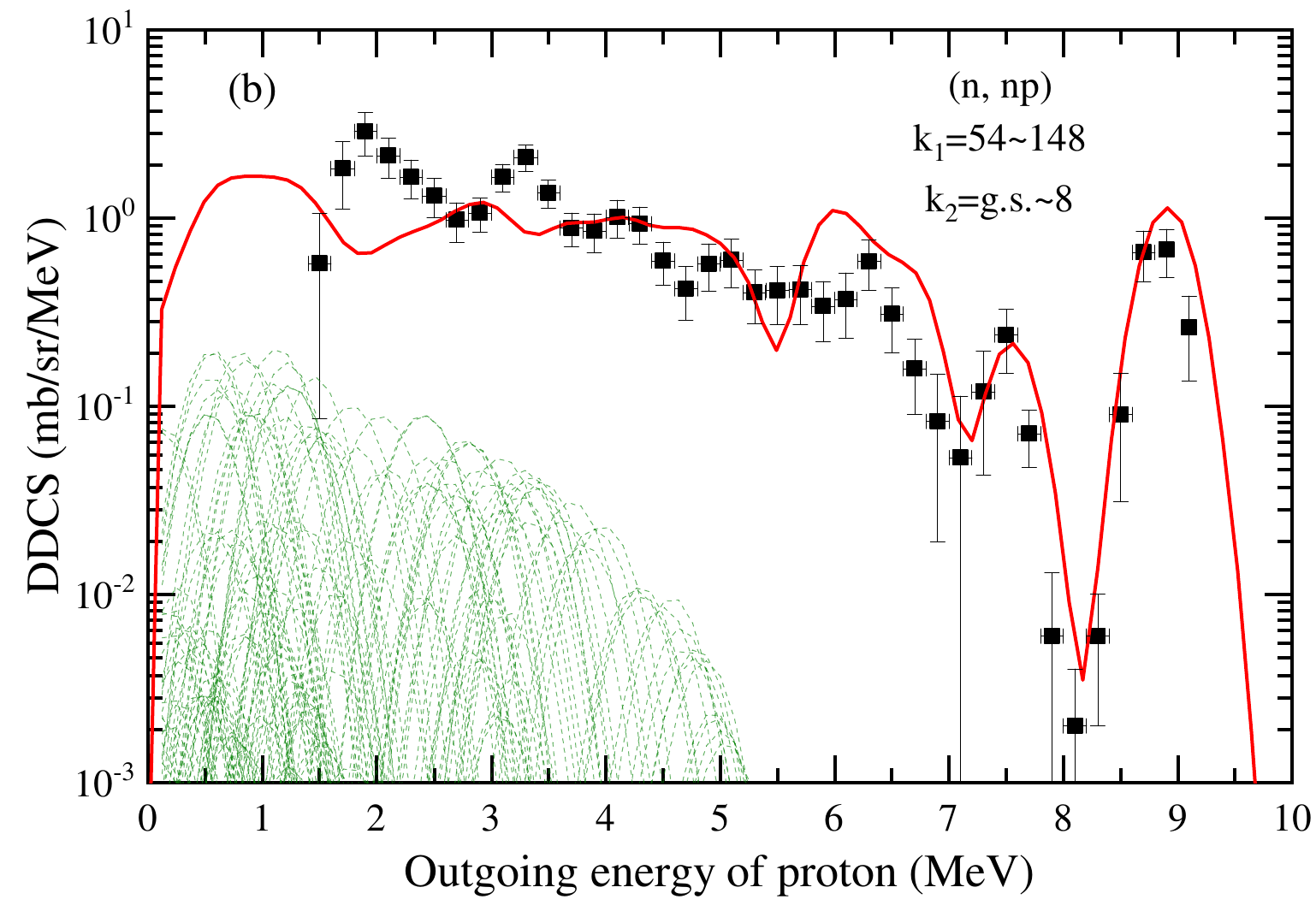}
	}
    \caption{(Color online) Partial and total DDCS of the outgoing proton for $n + ^{19}$F reaction with an outgoing angle $90^{\circ}$ at $E_n$ = 14.2 MeV in LS. The black points denote the experimental data taken from Ref. \cite{kondo2011measurement}, and the red solid lines denote the results of this work. The blue dashed lines denote the partial spectra of the first outgoing protons from the compound nucleus $^{20}\text{F}^*$ to the ground state up to 42th excited state of the first residual nucleus $^{19}\text{O}$ (a). The green dashed lines denote the partial spectra of the secondary emitted protons from the compound nucleus $^{20}\text{F}^*$, which firstly emits a neutron to the 54th-148th excited energy levels of first residual nucleus $^{19}\text{F}^*$, and then emits a proton to the ground state up to the 8th excited state of $^{18}\text{O}$ (b).}
    \label{Fig3}
\end{figure}
Taking the calculated DDCS of outgoing proton with outgoing angle $90^{\circ}$ at $E_n$ = 14.2 MeV as an example, the partial spectra are shown in Fig. \ref{Fig3}. The blue dashed lines denote the partial spectra of the first outgoing protons from the compound nucleus $^{20}\text{F}^*$ to the ground state up to the 42th excited state of the first residual nucleus $^{19}\text{O}$ (Fig. \ref{Fig3a}). It is evident that the contribution of the reaction channel (n, p) dominates the DDCS of outgoing proton. Each discrete peak experimentally observed originates from the superposition of two or more energy levels, reflecting the energy level structure of the residual nucleus $^{19}\text{O}$, including both its ground and excited states. The examination of peak positions and shapes offer significant insights into the nuclear energy level structure of $^{19}\text{O}$, as well as its dynamic behavior throughout the reaction process. The green dashed lines denote the partial spectra of the secondary emitted protons from the compound nucleus $^{20}\text{F}^*$, which firstly emits a neutron to the 54th-148th excited energy levels of first residual nucleus $^{19}\text{F}^*$, and then emits a proton to the ground state up to the 8th excited state of $^{18}\text{O}$. It is also of great importance to take into account the contribution of the secondary emitted protons, especially in the low energy region.

\begin{figure}[!htp]
    \centering
    \includegraphics[width=0.6\linewidth]{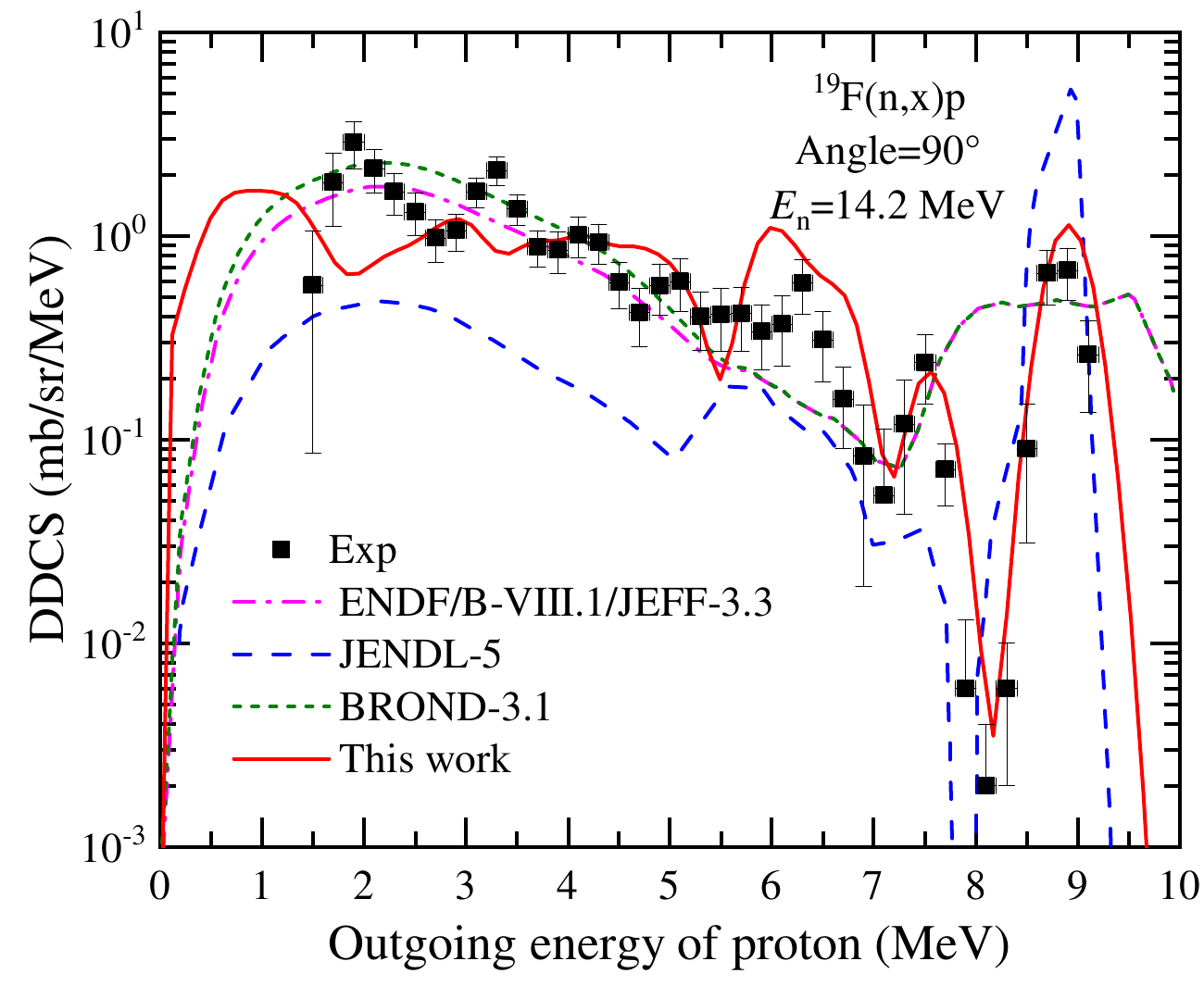}
    \caption{(Color online) Comparison of the DDCS of the emitted proton calculated in this work versus those recommended by major international databases for $n + ^{19}$F at $E_n$ = 14.2 MeV with $90^{\circ}$ angle. The black points denote the experimental data taken from Ref. \cite{kondo2011measurement}. The red solid, pink dash-dotted, blue dashed and green dotted lines denote the results of this work, ENDF/B-VIII.1, JENDL-5, and BROND-3.1, respectively.}
    \label{Fig4}
\end{figure}
Fig. 4 shows a comparison of the DDCS of outgoing proton calculated in this work versus those recommended by major international databases for $n + ^{19}$F at $E_n$ = 14.2 MeV with $90^{\circ}$ angle. The black points denote the experimental data taken from Ref. \cite{kondo2011measurement}. The red solid, pink dash-dotted, blue dashed and green dotted lines denote the results of this work, ENDF/B-VIII.1, JENDL-5, and BROND-3.1, respectively. The results evaluated by ENDF/B-VIII.1 and BROND-3.1 can partly reproduce the experimental DDCS in low energy region of outgoing protons. However, they fail to describe the discrete peaks in high outgoing energy region. This discrepancy primarily stems from the assumption of isotropic angular distributions \cite{ENDF-B-VIII.1,BROND-3.1}, which neglects the influence of the energy level structure of the residual nucleus $^{19}$O. The results evaluated by JENDL-5 show significant discrepancies with the experimental data.

\subsection{DDCS of outgoing deuteron }
\begin{figure}[!htp]
    \centering
    \includegraphics[width=0.6\hsize]{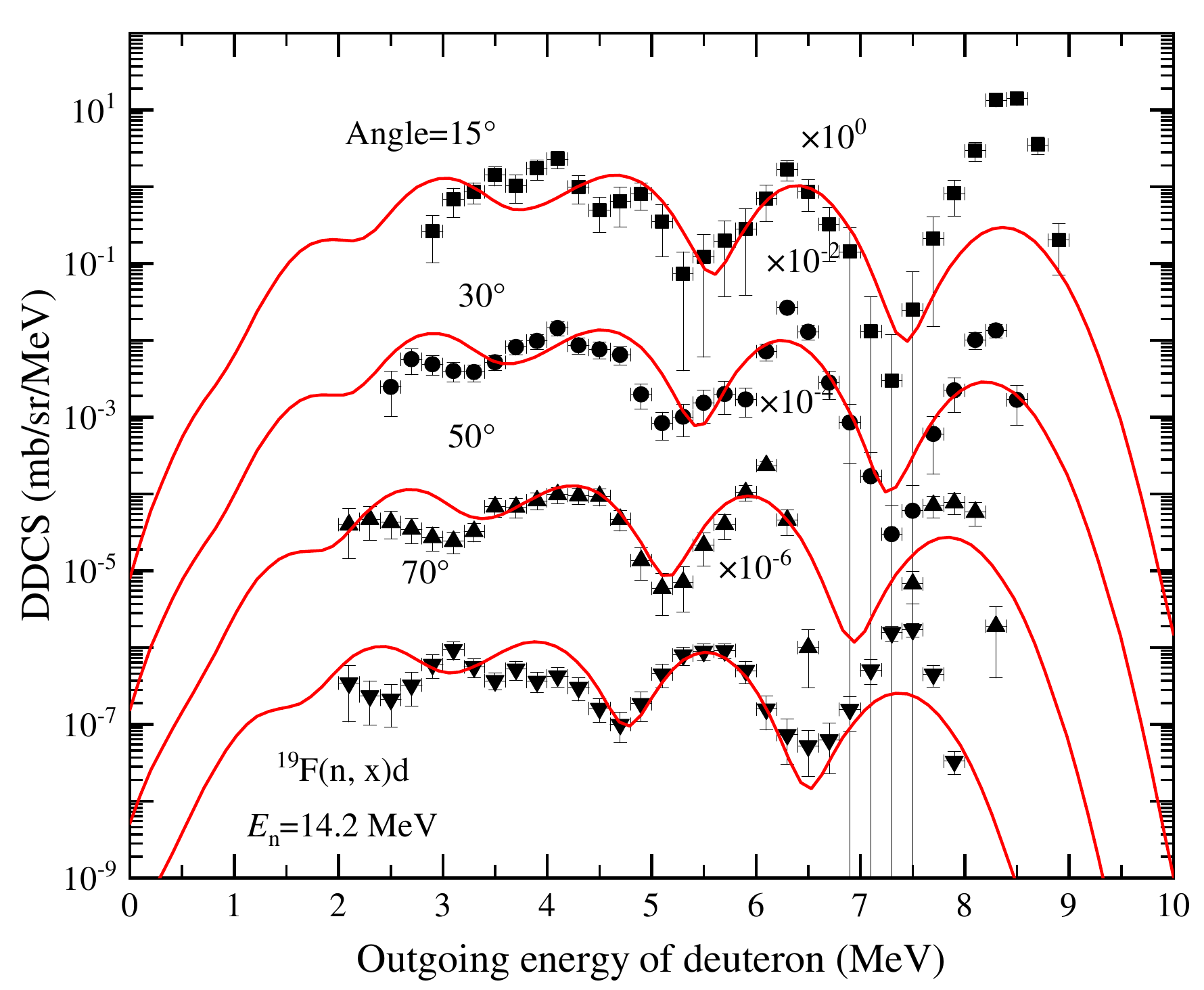}
    \caption{(Color online) Same as Fig. \ref{Fig1},  but for outgoing deuteron at outgoing angles $15^{\circ}$, $30^{\circ}$, $50^{\circ}$, and $70^{\circ}$.}
    \label{Fig5}
\end{figure}
\begin{figure}[!htp]
    \centering
    \includegraphics[width=0.6\hsize]{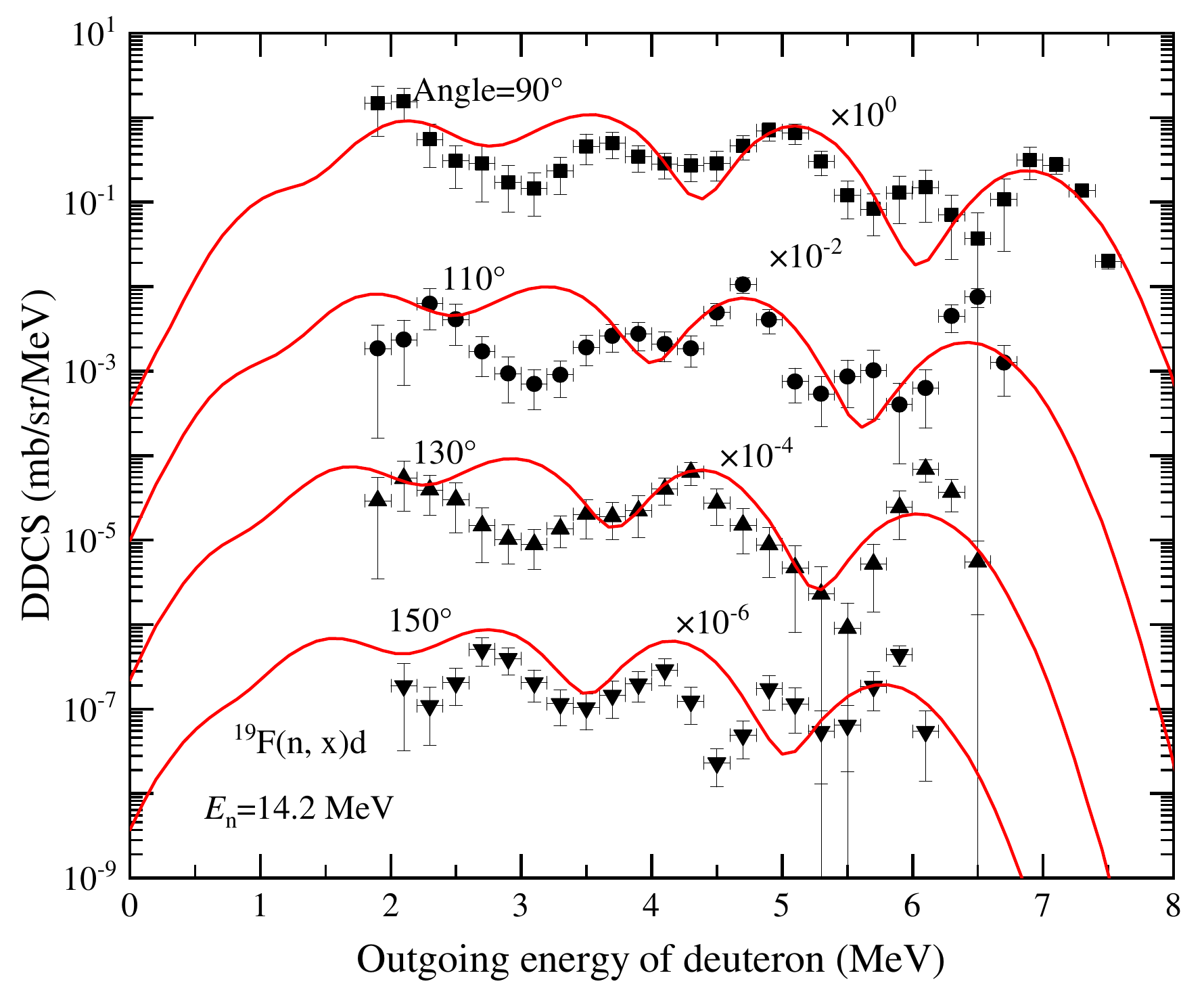}
    \caption{(Color online) Same as Fig. \ref{Fig1}, but for outgoing deuteron at outgoing angles $90^{\circ}$, $110^{\circ}$, $130^{\circ}$, and $150^{\circ}$.}
    \label{Fig6}
\end{figure}
\begin{figure}[!htp]
    \centering
    \includegraphics[width=0.6\hsize]{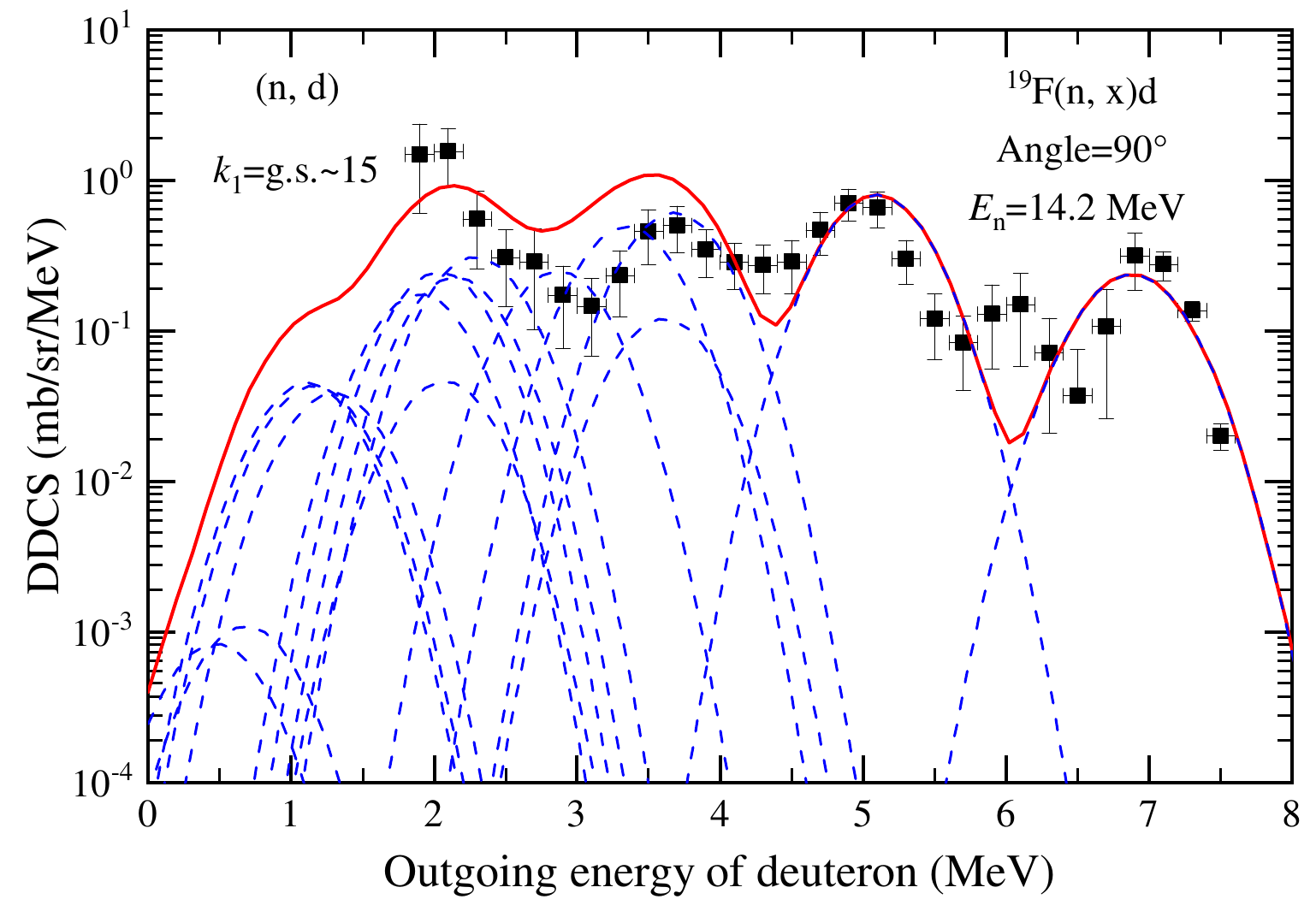}
    \caption{(Color online) Partial and total DDCS of the outgoing deuteron for $\text{n} + ^{19}\text{F}$ reaction with $90^{\circ}$ outgoing angle at $E_n$ = 14.2 MeV in LS. The black points denote the experimental data taken from Ref. \cite{kondo2011measurement}, and the red solid line denotes the result of this work. The blue dashed lines denote the partial spectra of the first outgoing deuteron from the compound nucleus $^{20}\text{F}^*$ to the ground state up to 15th excited state of the first residual nucleus $^{18}\text{O}$.}
    \label{Fig7}
\end{figure}
The comparisons of the calculated total double-differential cross sections of the outgoing deuteron with the measured data are shown in Figs. \ref{Fig5} and \ref{Fig6} at $E_n$=14.2 MeV for outgoing angles of $15^{\circ}$, $30^{\circ}$, $50^{\circ}$, $70^{\circ}$, $90^{\circ}$, $110^{\circ}$, $130^{\circ}$ and $150^{\circ}$, respectively.  The black points represent the experimental data taken from Ref. \cite{kondo2011measurement}, and the red solid lines represent the results of this work. Similarly, the calculated results agree well with the experimental DDCS of the outgoing deuteron, except at the outgoing angle 15$^\circ$. The theoretical calculations are obviously lower than the experimental data at the outgoing angle 15$^\circ$ angle in the 8–9 MeV outgoing energy region. This reason is same as for the outgoing proton mentioned above. In addtion, the calculated results exhibit pronounced peaks at outgoing deuteron energy around 3 MeV only for outgoing angles $110^\circ$ and $130^\circ$, which are completely opposite to the experimental data. This discrepancy may have arisen from some unknown causes during the measurement process, and it is hoped that experimental physicists can further verify such data.

\begin{figure}[!htp]
    \centering
    \includegraphics[width=0.6\hsize]{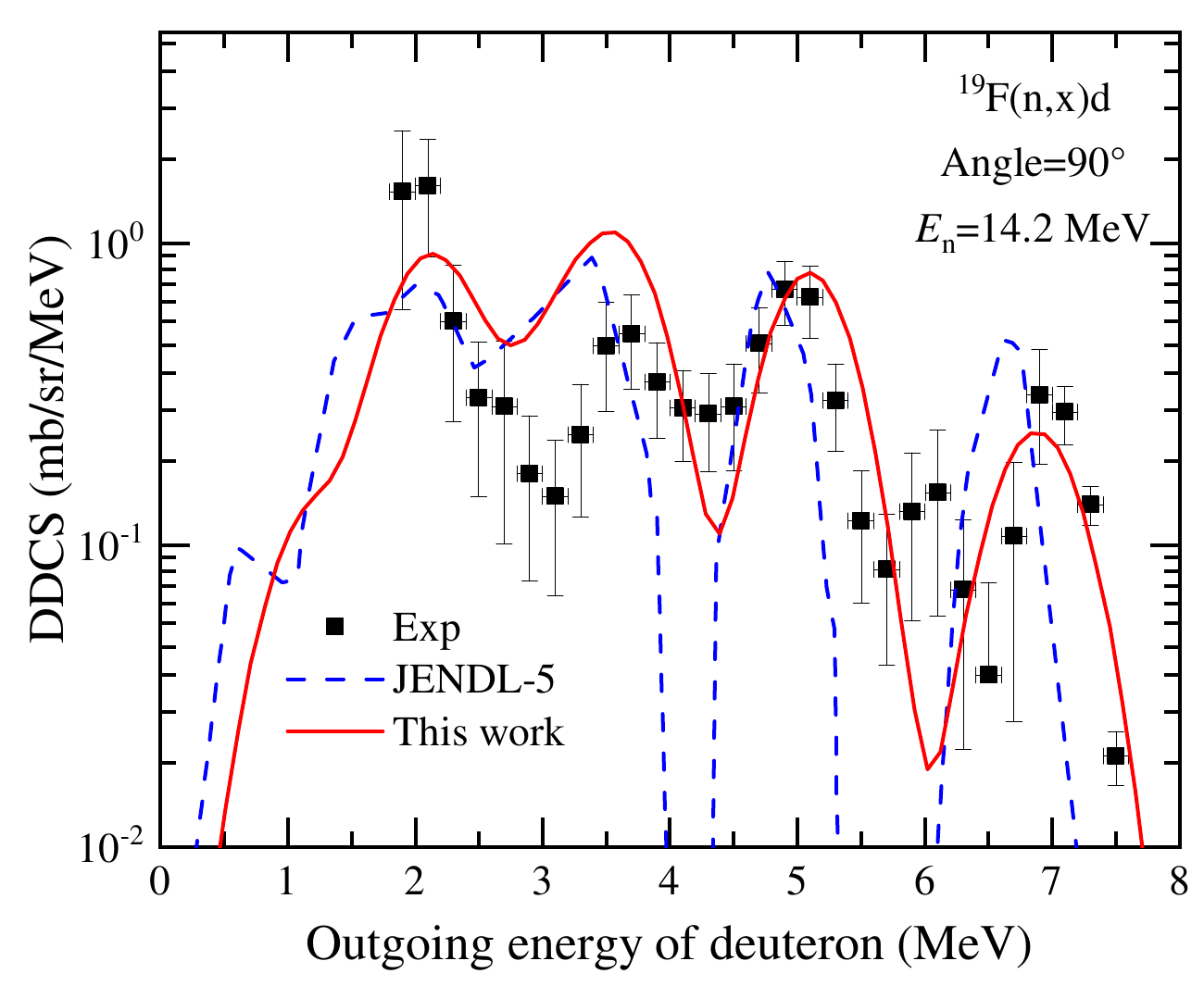}
    \caption{(Color online) Comparison of the calculated DDCS of outgoing deuteron for $n + ^{19}$F reaction with the experimental data and evaluated results at $E_n$ = 14.2 MeV with outgoing angle $90^{\circ}$. The black points denote the experimental data taken from Ref. \cite{kondo2011measurement}. The red solid and blue dashed lines denote the results of this work and JENDL-5, respectively.}
    \label{Fig8}
\end{figure}

Fig. \ref{Fig7} shows the partial and total DDCS of the outgoing deuteron for $n + ^{19}$F reaction with an outgoing angle $90^{\circ}$ at $E_n$ = 14.2 MeV in LS. The blue dashed lines denote the partial spectra of the first outgoing deuteron from the compound nucleus $^{20}\text{F}^*$ to the ground state up to 15th excited state of the first residual nucleus $^{18}\text{O}$. These energy spectra from $(n, d)$ channel collectively contribute to the total DDCS of outgoing deuteron. Each discrete peak experimentally observed reflects the influence of the energy level structure of the residual nucleus $^{18}$O.

In Fig. \ref{Fig8}, the calculated total DDCS of the outgoing deuteron are compared with the experimental data and the evaluated data from JENDL-5. The black points denote the experimental data taken from Ref. \cite{kondo2011measurement}. The red solid and blue dashed lines denote the results of this work and JENDL-5, respectively. The calculated results in this paper and the evaluated data from JENDL-5 both reasonably reproduce the measured data, but the former aligns more closely with the experimental data than the latter.

\subsection{DDCS of outgoing triton}
\begin{figure}[!ht]
    \centering
    \includegraphics[width=0.6\hsize]{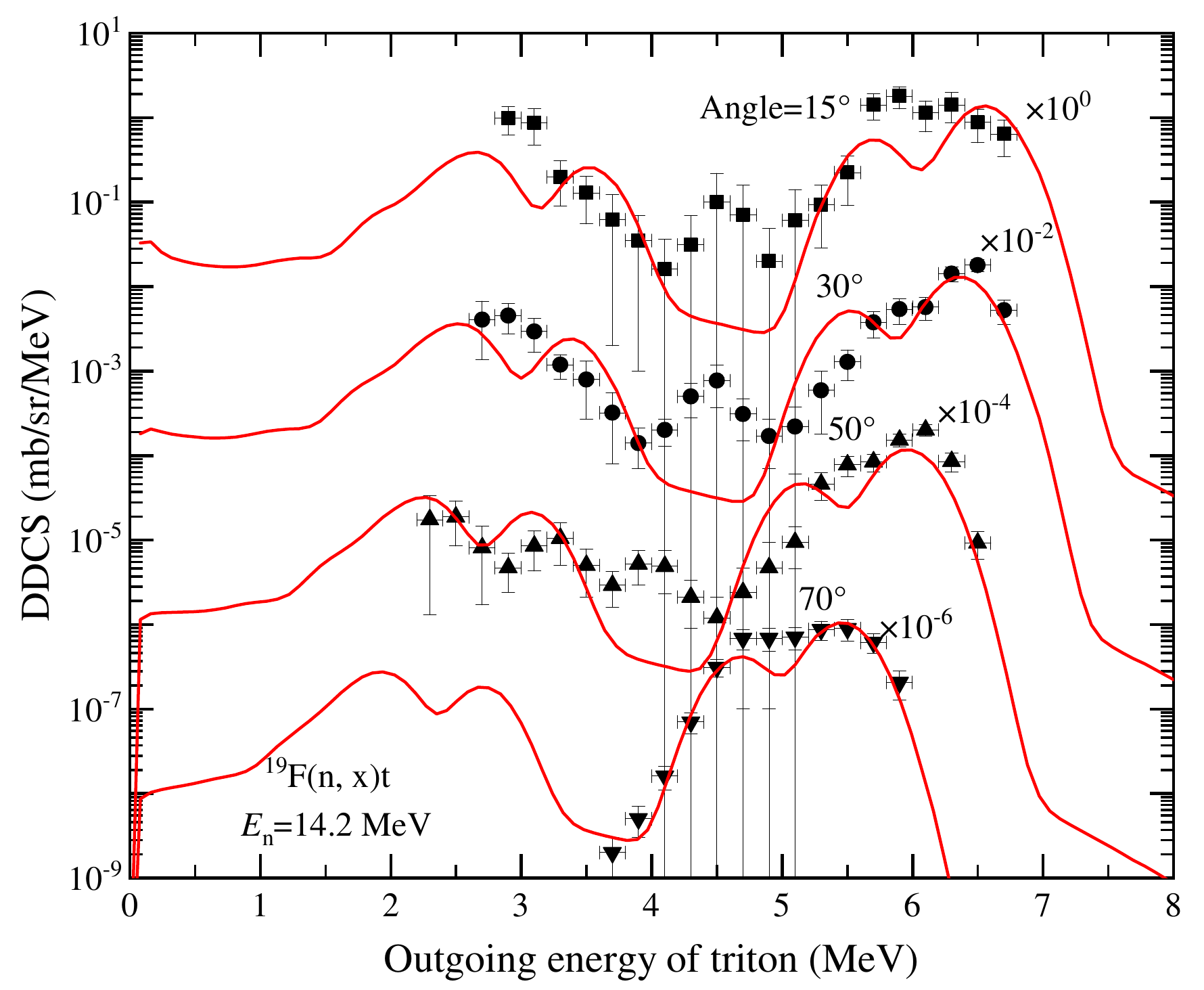}
    \caption{(Color online) Same as Fig.~\ref{Fig1},  but for outgoing triton at outgoing angles $15^{\circ}$, $30^{\circ}$, $50^{\circ}$, and $70^{\circ}$.}
    \label{Fig9}
\end{figure}
\begin{figure}[!ht]
    \centering
    \includegraphics[width=0.6\hsize]{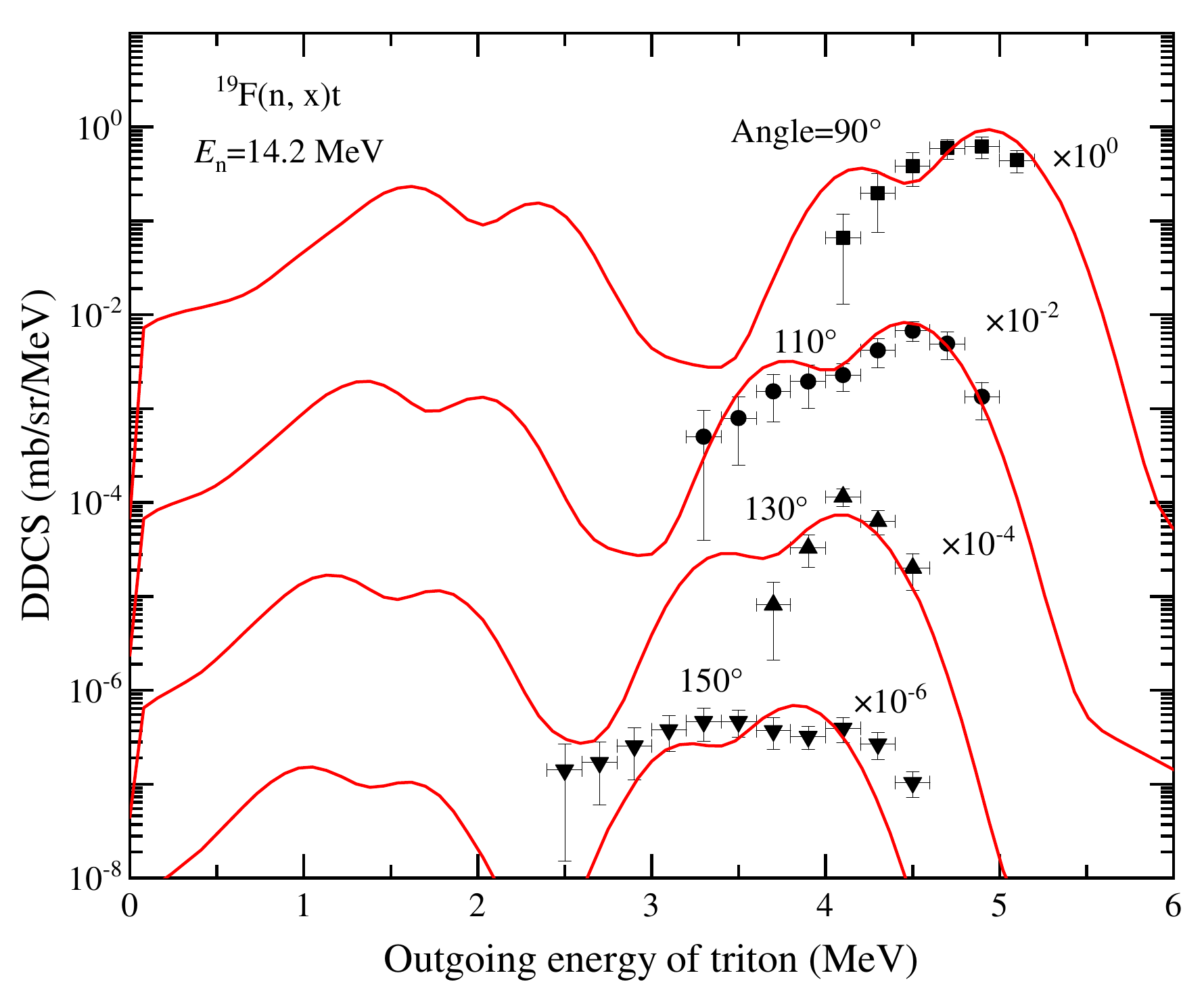}
    \caption{(Color online)  Same as Fig.~\ref{Fig1},  but for outgoing triton at outgoing angles $90^{\circ}$, $110^{\circ}$, $130^{\circ}$, and $150^{\circ}$.}
    \label{Fig10}
\end{figure}
\begin{figure}[!htb]
\centering
    \subfigure{
		\label{Fig11a}
		\includegraphics[width=0.45\hsize]{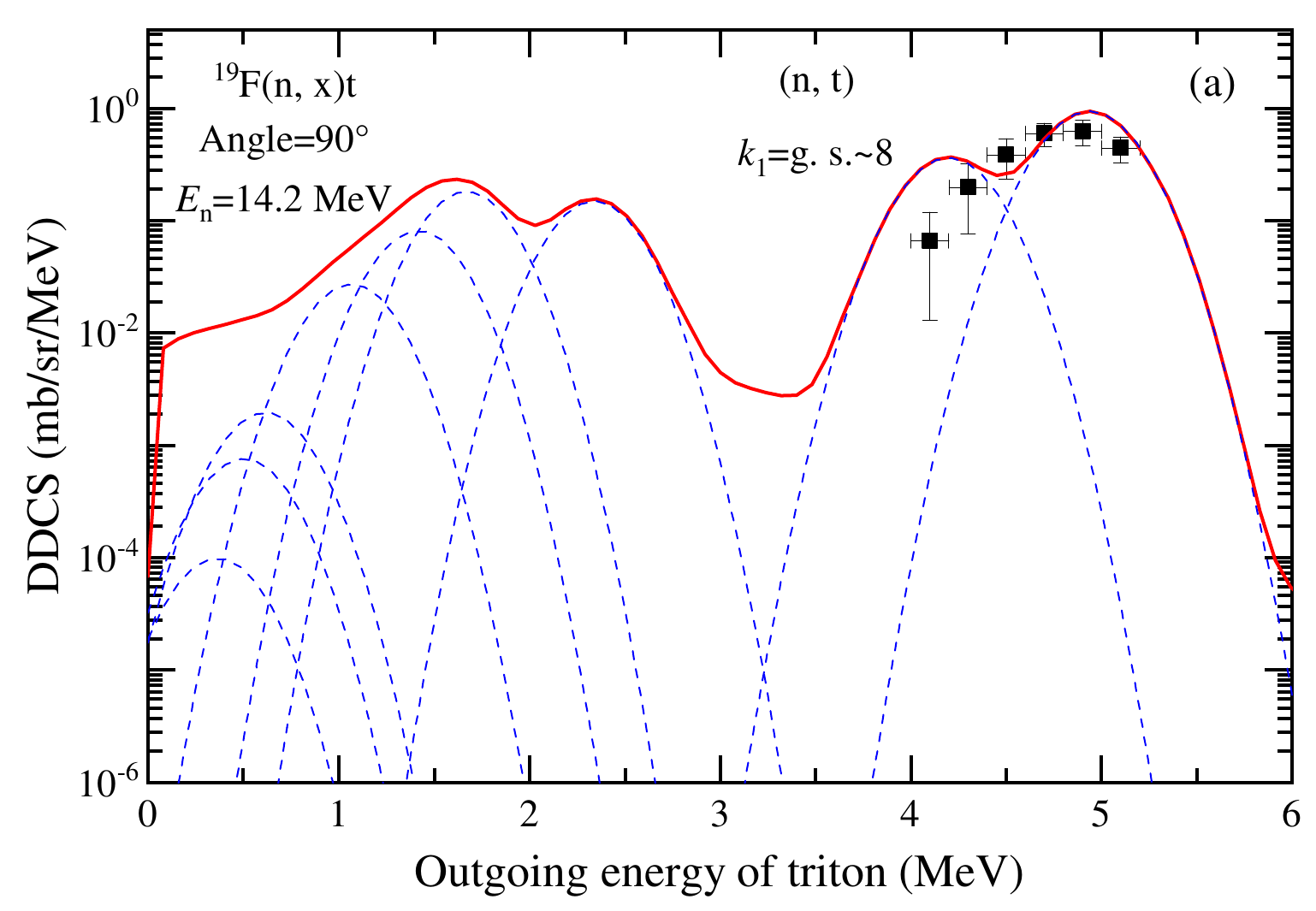}
	}
    \subfigure{
		\label{Fig11b}
		\includegraphics[width=0.45\hsize]{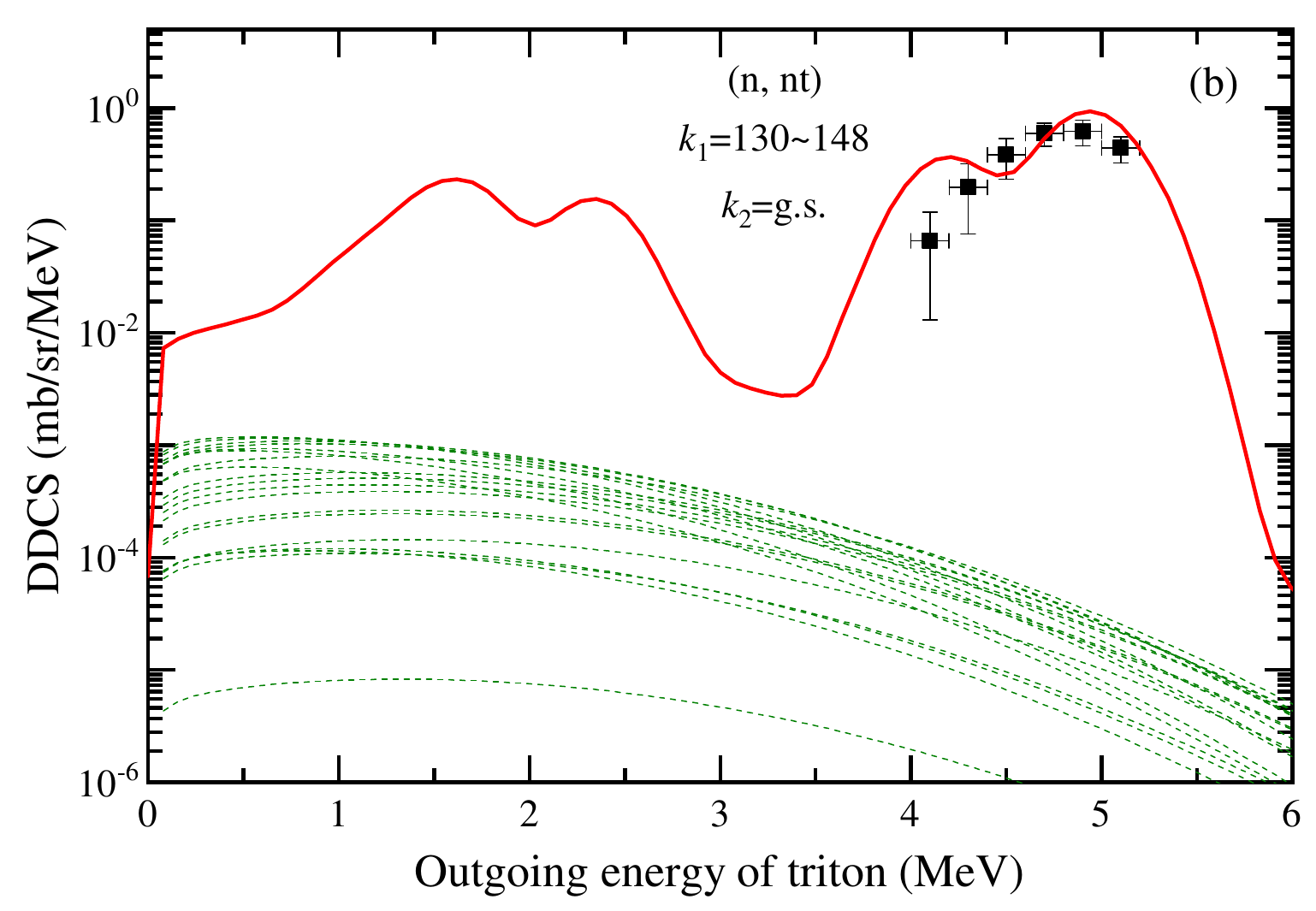}
	}
    \caption{(Color online) Partial and total DDCS of the outgoing triton for $n + ^{19}$F reaction with an outgoing angle $90^{\circ}$ at $E_n$ = 14.2 MeV in LS. The black points denote the experimental data taken from Ref. \cite{kondo2011measurement}, and the red solid lines denote the results of this work. The blue dashed lines denote the partial spectra of the first outgoing tritons from the compound nucleus $^{20}\text{F}^*$ to the ground state up to 8th excited state of the first residual nucleus $^{17}\text{O}$ (a). The green dashed lines denote the partial spectra of the secondary emitted tritons from the compound nucleus $^{20}\text{F}^*$, which first emits a neutron to the 130th-148th excited energy levels of first residual nucleus $^{19}\text{F}$, and then emits a triton to the ground state of $^{16}\text{O}$  (b).}
    \label{Fig11}
\end{figure}
The comparisons of the calculated total DDCS of outgoing triton with the measured data are shown in Figs. \ref{Fig9} and \ref{Fig10} at the incident neutron energy $E_n$=14.2 MeV for outgoing angles of $15^{\circ}$, $30^{\circ}$, $50^{\circ}$, $70^{\circ}$, $90^{\circ}$, $110^{\circ}$, $130^{\circ}$ and $150^{\circ}$, respectively. The black points represent the experimental data taken from Ref. \cite{kondo2011measurement}, and the red solid lines represent the results of this work. One can see that the calculated results are in good agreement with the experimental data. However, there is an unknown peak around 4-4.5 MeV as shown in Fig. \ref{Fig9} at the outgoing angles of $15^{\circ}$, $30^{\circ}$, and $50^{\circ}$, respectively. The energy of the particle corresponds to an excited state of $^{17}\text{O}$ at approximately 2 MeV, but it is not likely to be a real new excited state in $^{17}\text{O}$, because there is no corresponding state in the mirror nucleus $^{17}\text{F}$ \cite{tilley1993energy}. It is still unclear what kind of mechanism contributes to the formation of this peak. It is expected that these unidentified peaks can be further confirmed through future experimental investigations.

\begin{figure}[!ht]
    \centering
    \includegraphics[width=0.6\linewidth]{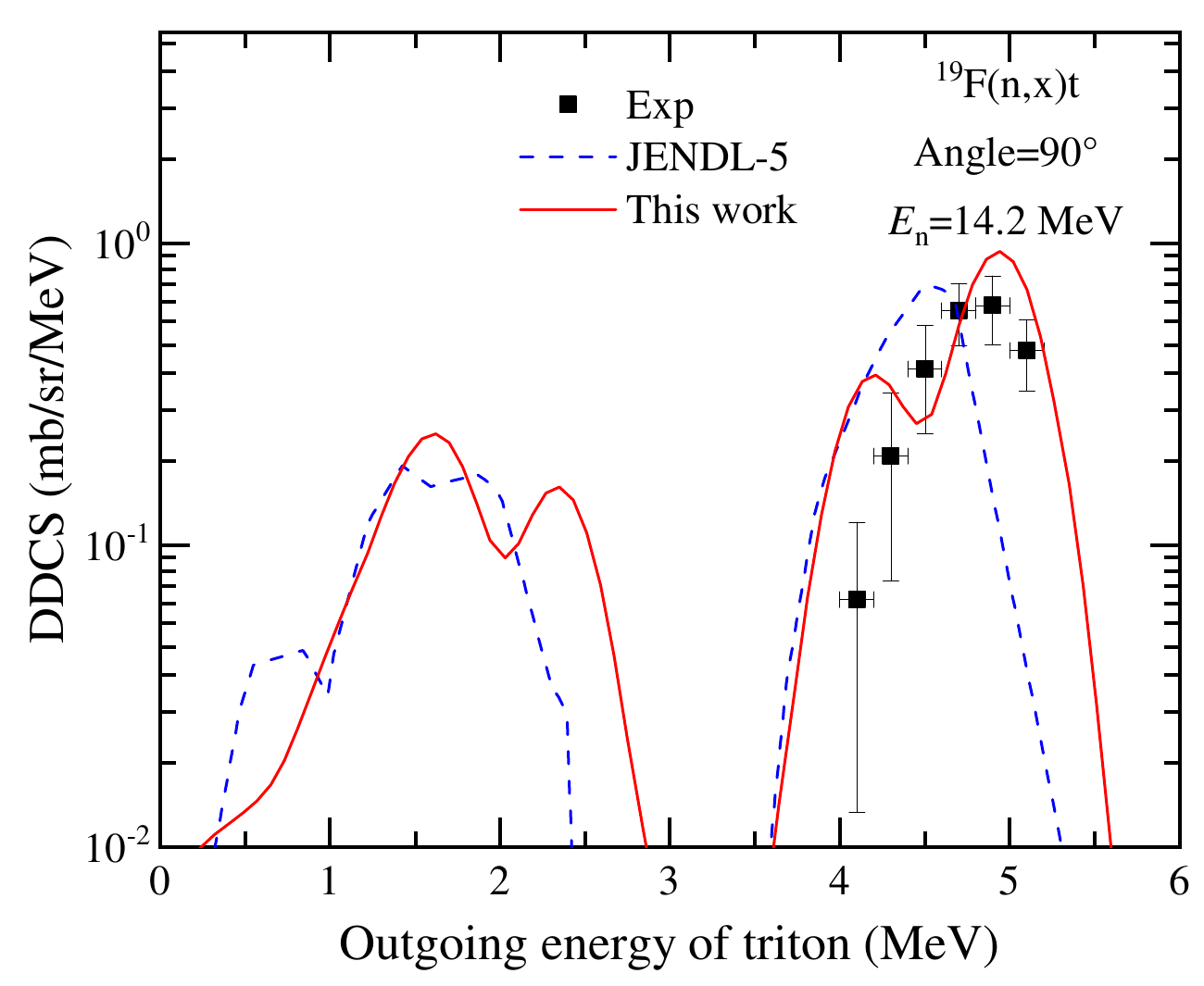}
    \caption{(Color online) Comparison of the calculated DDCS of outgoing triton at $90^{\circ}$ angl for $n + ^{19}$F reaction with the experimental data and evaluated results at $E_n$ = 14.2 MeV. The black points denote the experimental data taken from Ref. \cite{kondo2011measurement}. The red solid and blue dashed lines denote the results of this work and JENDL-5, respectively.}
    \label{Fig12}
\end{figure}
The partial and total DDCS of outgoing triton for $n + ^{19}$F reaction are shown in Fig. \ref{Fig11} at $E_n$ = 14.2 MeV with $90^{\circ}$ outgoing angle. The blue dashed lines denote the partial spectra of the first outgoing tritons from the compound nucleus $^{20}\text{F}^*$ to the ground state up to 8th excited state of the first residual nucleus $^{17}\text{O}$ (Fig. \ref{Fig11a}). Each discrete peak matchs an energy level of $^{17}$O. The green dashed lines denote the partial spectra of the secondary emitted tritons from the compound nucleus $^{20}\text{F}^*$, which first emits a neutron to the 130th-148th excited energy levels of the first residual nucleus $^{19}\text{F}$, and then emits a triton to the ground state of $^{16}\text{O}$  (Fig. \ref{Fig11b}). Here, $k_2$ denotes the energy levels of the secondary residual nuclei. Although reaction channel $(n, nt)$ contributes less significantly to the overall spectra compared to $(n, t)$ channel, its influence extends over a broader range of outgoing energies.

Fig. \ref{Fig12} shows a comparison of the DDCS of outgoing triton with the experimental data and evaluated results at $E_n$ = 14.2 MeV with $90^{\circ}$ angle for $n + ^{19}$F reaction. The black points denote the experimental data taken from Ref. \cite{kondo2011measurement}. The red solid and blue dashed lines denote the results of this work and JENDL-5, respectively. The calculated results in this paper and the evaluated data from JENDL-5 both reasonably reproduce the measured data, but the former aligns more closely with the experimental data than the latter.

\subsection{DDCS of outgoing alpha }
\begin{figure}[!ht]
    \centering
    \includegraphics[width=0.6\hsize]{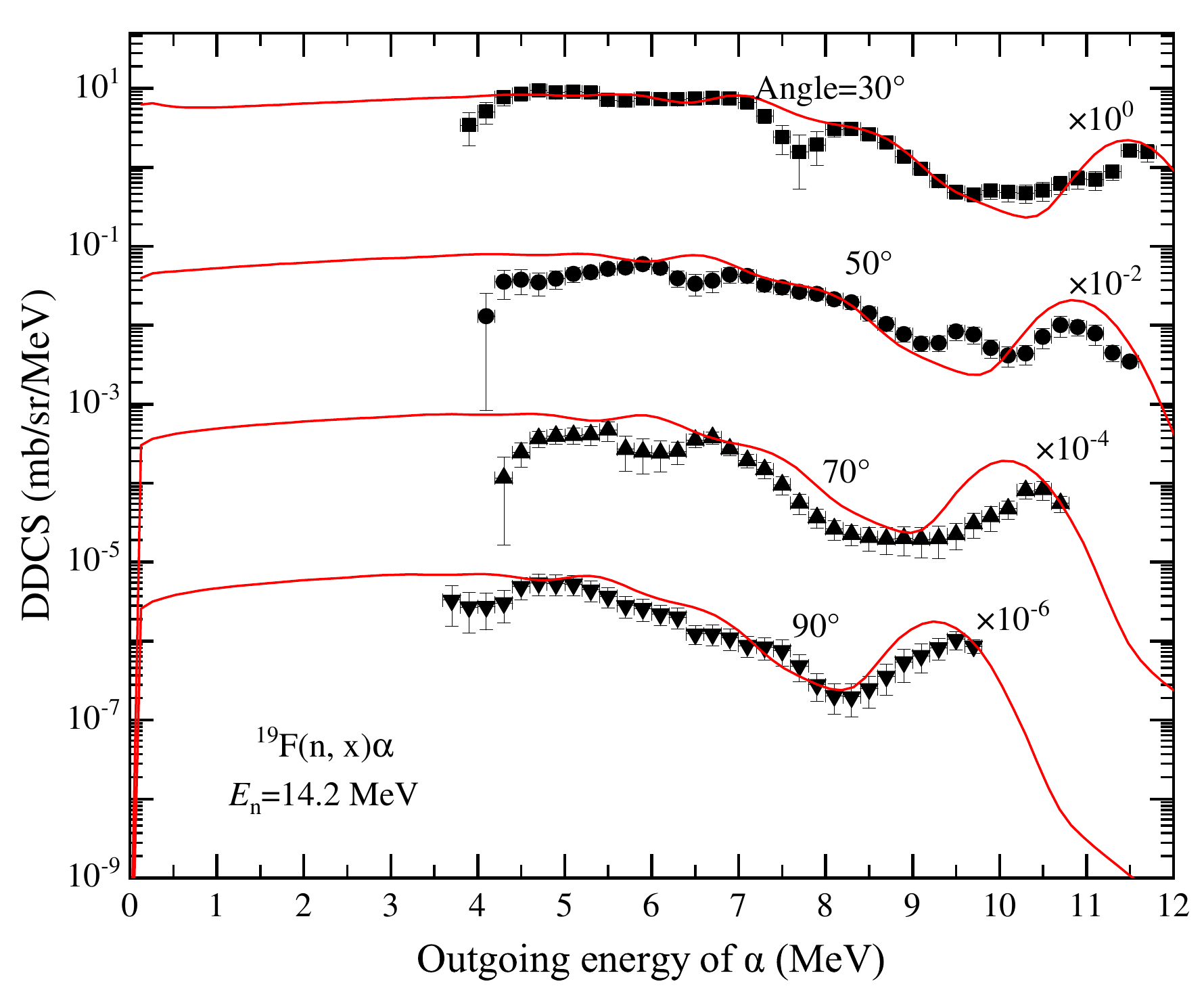}
    \caption{(Color online)  Same as Fig. \ref{Fig1},  but for outgoing $\alpha$ at outgoing angles $30^{\circ}$, $50^{\circ}$, $70^{\circ}$, and $90^{\circ}$.}
    \label{Fig13}
\end{figure}
\begin{figure}[!ht]
    \centering
    \includegraphics[width=0.6\hsize]{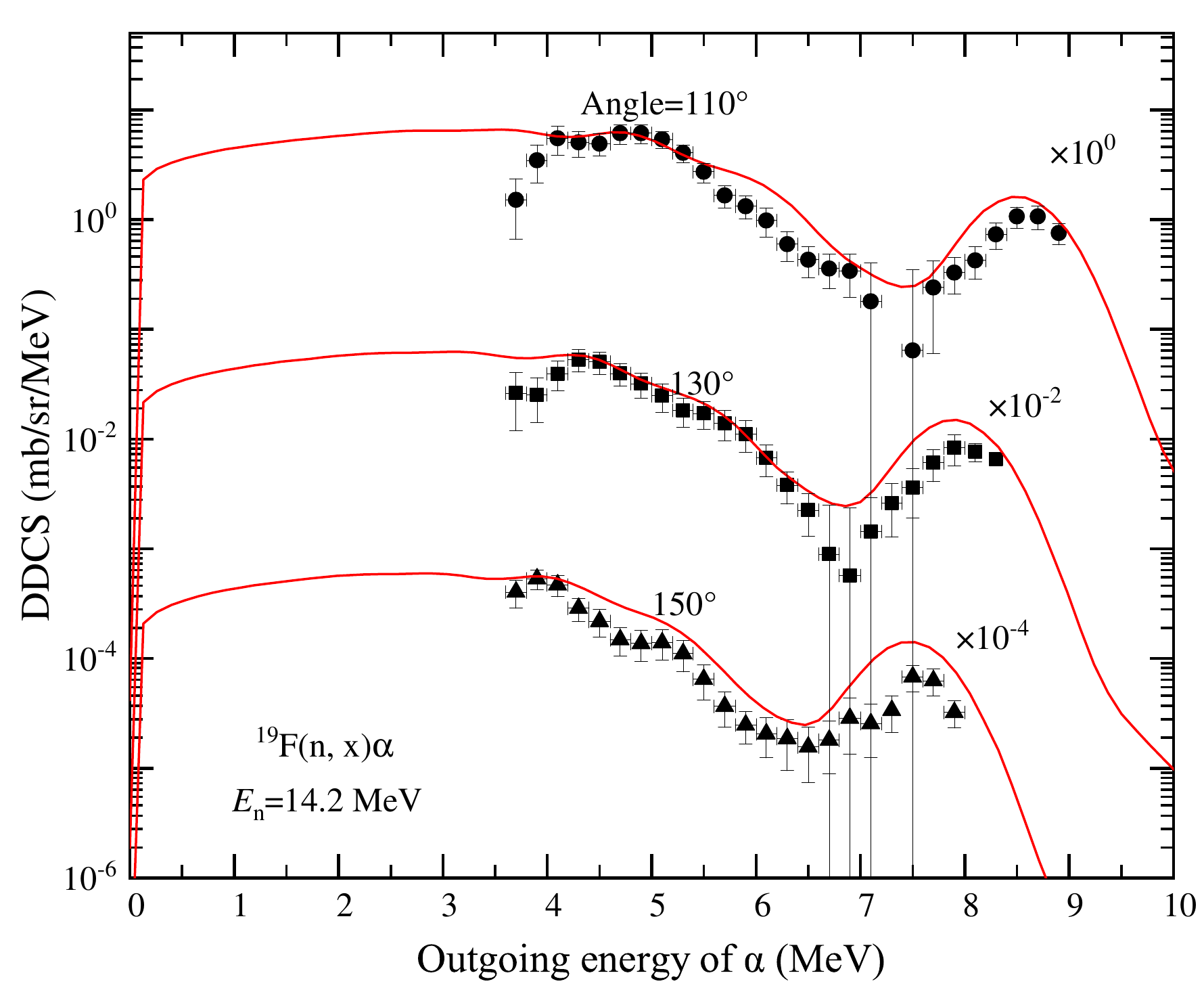}
    \caption{(Color online) Same as Fig. \ref{Fig1},  but for outgoing $\alpha$ at outgoing angles  $110^{\circ}$, $130^{\circ}$, and $150^{\circ}$.}
    \label{Fig14}
\end{figure}
The comparisons of the calculated double-differential cross sections of the total outgoing $\alpha$ particle with the measured data are shown in Figs. \ref{Fig13} and \ref{Fig14} at the incident neutron energy $E_n$=14.2 MeV with outgoing angles of $30^{\circ}$, $50^{\circ}$, $70^{\circ}$, $90^{\circ}$, $110^{\circ}$, $130^{\circ}$ and $150^{\circ}$, respectively. One can see that the calculated results agree well with the experimental data.

\begin{figure}[!htb]
\centering
    \subfigure{
		\label{Fig15a}
		\includegraphics[width=0.45\hsize]{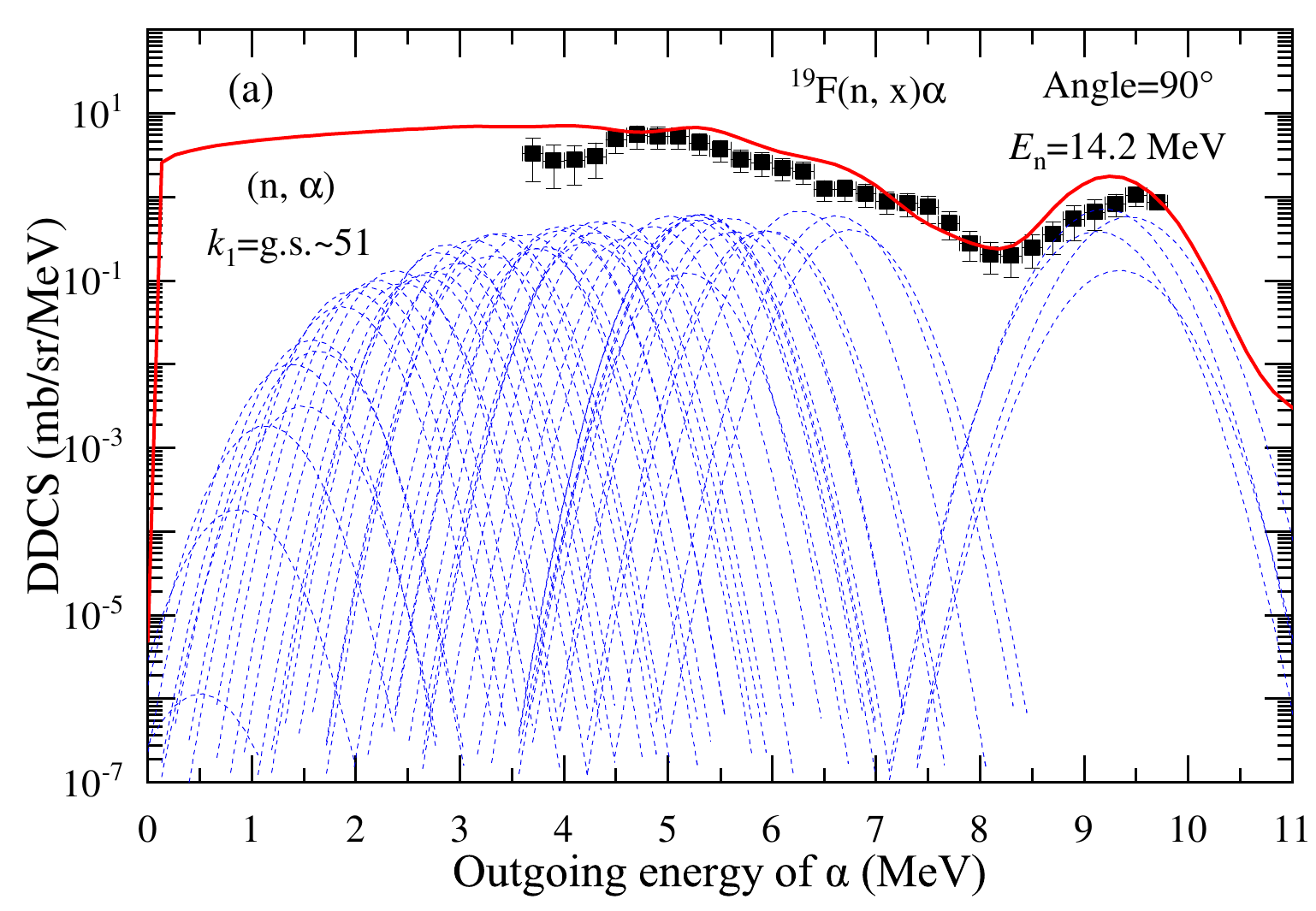}
	}
    \subfigure{
		\label{Fig15b}
		\includegraphics[width=0.45\hsize]{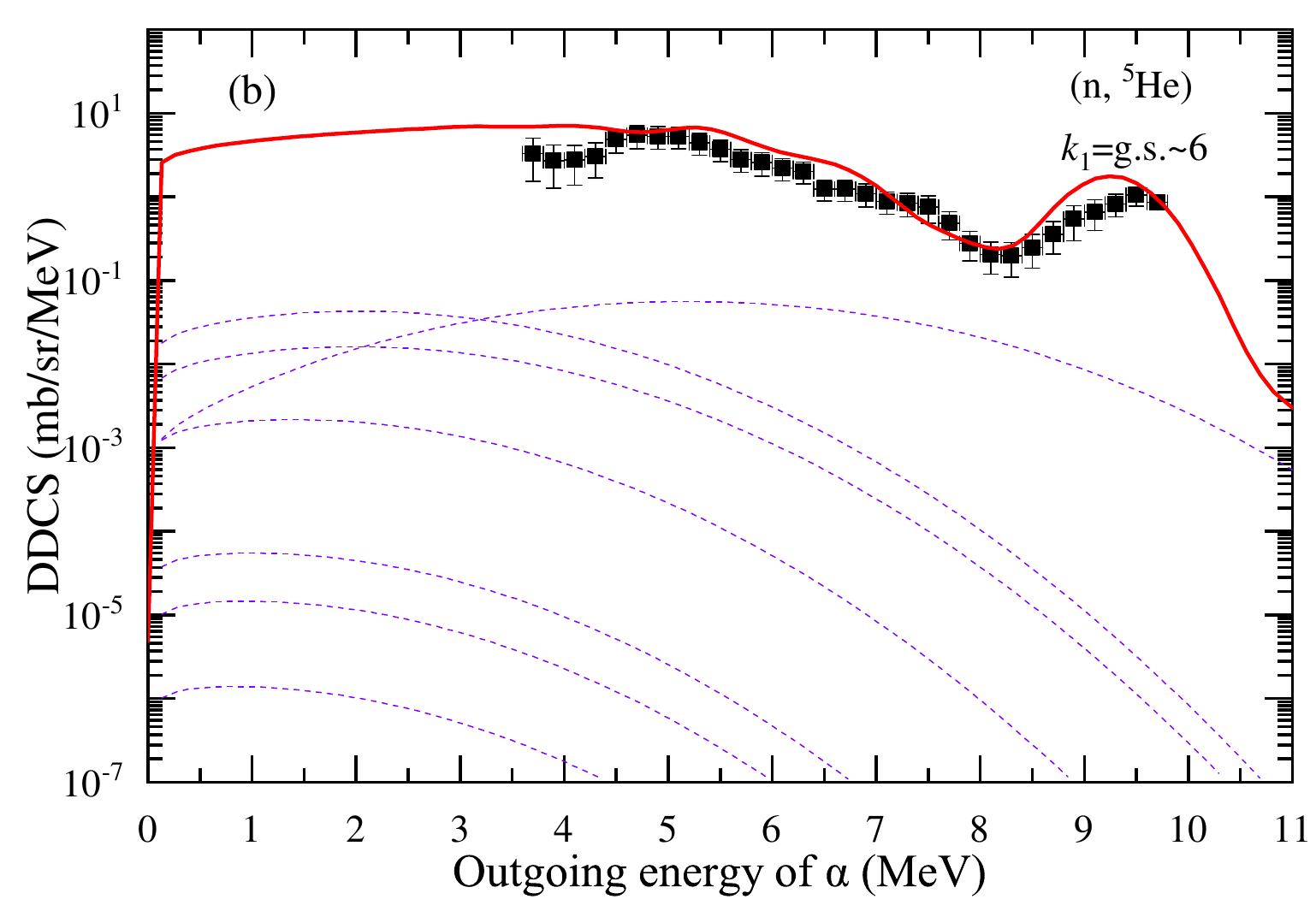}
	}
    \subfigure{
		\label{Fig15c}
		\includegraphics[width=0.45\hsize]{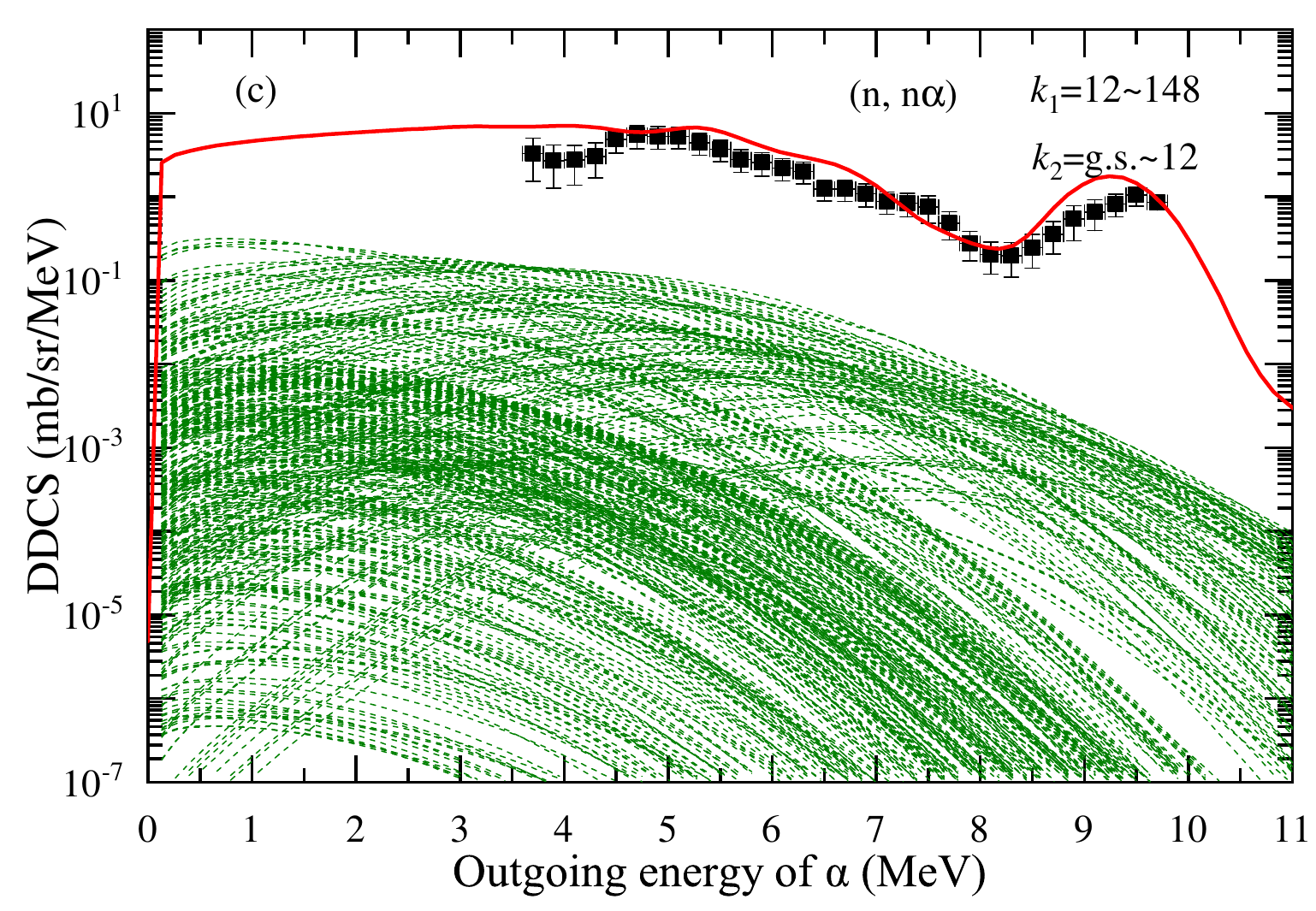}
	}
    \subfigure{
		\label{Fig15d}
		\includegraphics[width=0.45\hsize]{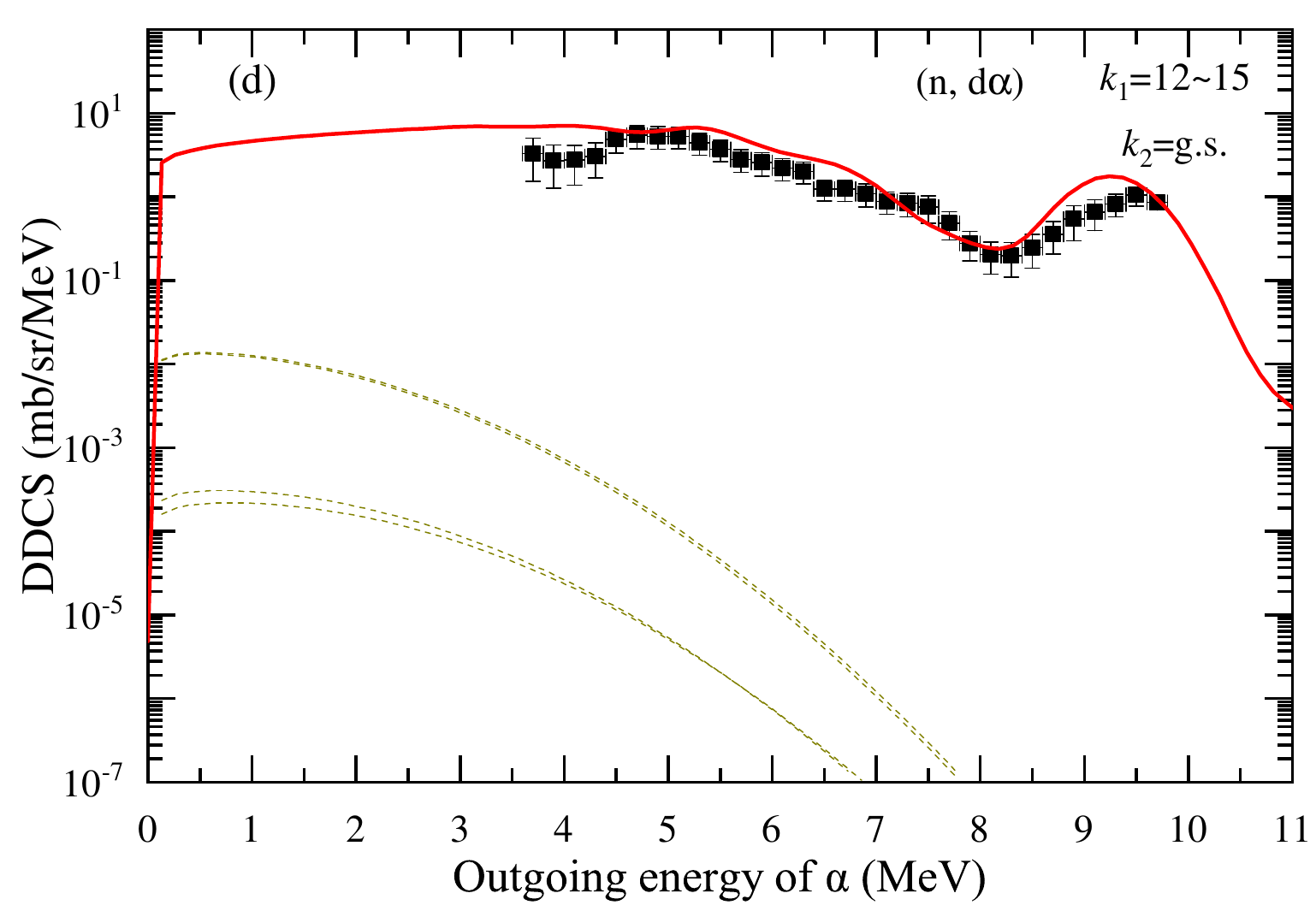}
	}
    \caption{(Color online) Partial and total DDCS of the outgoing $\alpha$ for $n + ^{19}$F reaction with an outgoing angle $90^{\circ}$ at $E_n$ = 14.2 MeV in LS. The black points denote the experimental data taken from Ref. \cite{kondo2011measurement}, and the red solid lines denote the results of this work. The blue dashed lines denote the partial spectra of the first outgoing $\alpha$ from the compound nucleus $^{20}\text{F}^*$ to the ground state up to 51th excited state of the first residual nucleus $^{16}\text{N}$ (a). The purple dashed lines denote the partial spectra of the first outgoing $^5\text{He}$ (which can breakup into $n$ and $\alpha$) from the compound nucleus $^{20}\text{F}^*$ to the ground state up to 6th excited state of the first residual nucleus $^{15}\text{N}$ (b). The green dashed lines denote the partial spectra of the secondary emitted $\alpha$ from the compound nucleus $^{20}\text{F}^*$, which first emits a neutron to the 12th-148th excited energy levels of first residual nucleus $^{19}\text{F}$, and then emits an $\alpha$ to the ground state up to the 12th excited state of $^{15}\text{N}$ (c). The yellow dashed lines denote the partial spectra of the secondary emitted $\alpha$ from the compound nucleus $^{20}\text{F}^*$, which first emits a deuteron to the 12th-15th excited energy levels of first residual nucleus $^{18}\text{O}$, and then emits an $\alpha$ to the ground state of $^{14}\text{C}$ (d).}
    \label{Fig15}
\end{figure}
Fig. \ref{Fig15} shows the partial and total double-differential cross-sections of the outgoing $\alpha$ particle for $\text{n} + ^{19}\text{F}$ reaction with $90^{\circ}$ outgoing angle at $E_n$ = 14.2 MeV in LS. The black points denote the experimental data taken from Ref. \cite{kondo2011measurement}, and the red solid lines denote the results of this work. The blue dashed lines denote the partial spectra of the first outgoing $\alpha$ from the compound nucleus $^{20}\text{F}^*$ to the ground state up to 51th excited state of the first residual nucleus $^{16}\text{N}$ (Fig. \ref{Fig15a}). Each experimentally observed discrete peak reflects the influence of the energy level structure of the residual nucleus $^{16}\text{N}$. The purple dashed lines denote the partial spectra of the first outgoing $^5\text{He}$ (which can breakup into n and $\alpha$) from the compound nucleus $^{20}\text{F}^*$ to the ground state up to 6th excited state of the first residual nucleus $^{15}\text{N}$ (Fig. \ref{Fig15b}). The green dashed lines denote the partial spectra of the secondary emitted $\alpha$ from the compound nucleus $^{20}\text{F}^*$, which first emits a neutron to the 12th-148th excited energy levels of first residual nucleus $^{19}\text{F}$, and then emits an $\alpha$ to the ground state up to the 12th excited state of $^{15}\text{N}$ (Fig. \ref{Fig15c}). The yellow dashed lines denote the partial spectra of the secondary emitted $\alpha$ from the compound nucleus $^{20}\text{F}^*$, which first emits a deuteron to the 12th-15th excited energy levels of first residual nucleus $^{18}\text{O}$, and then emits an $\alpha$ to the ground state of $^{14}\text{C}$ (Fig. \ref{Fig15d}). It is obvious that there are significant contributions from secondary particle emission to the double-differential cross sections of outgoing $\alpha$ particle for n + $^{19}$F reaction.

\begin{figure}[!ht]
    \centering
    \includegraphics[width=0.6\linewidth]{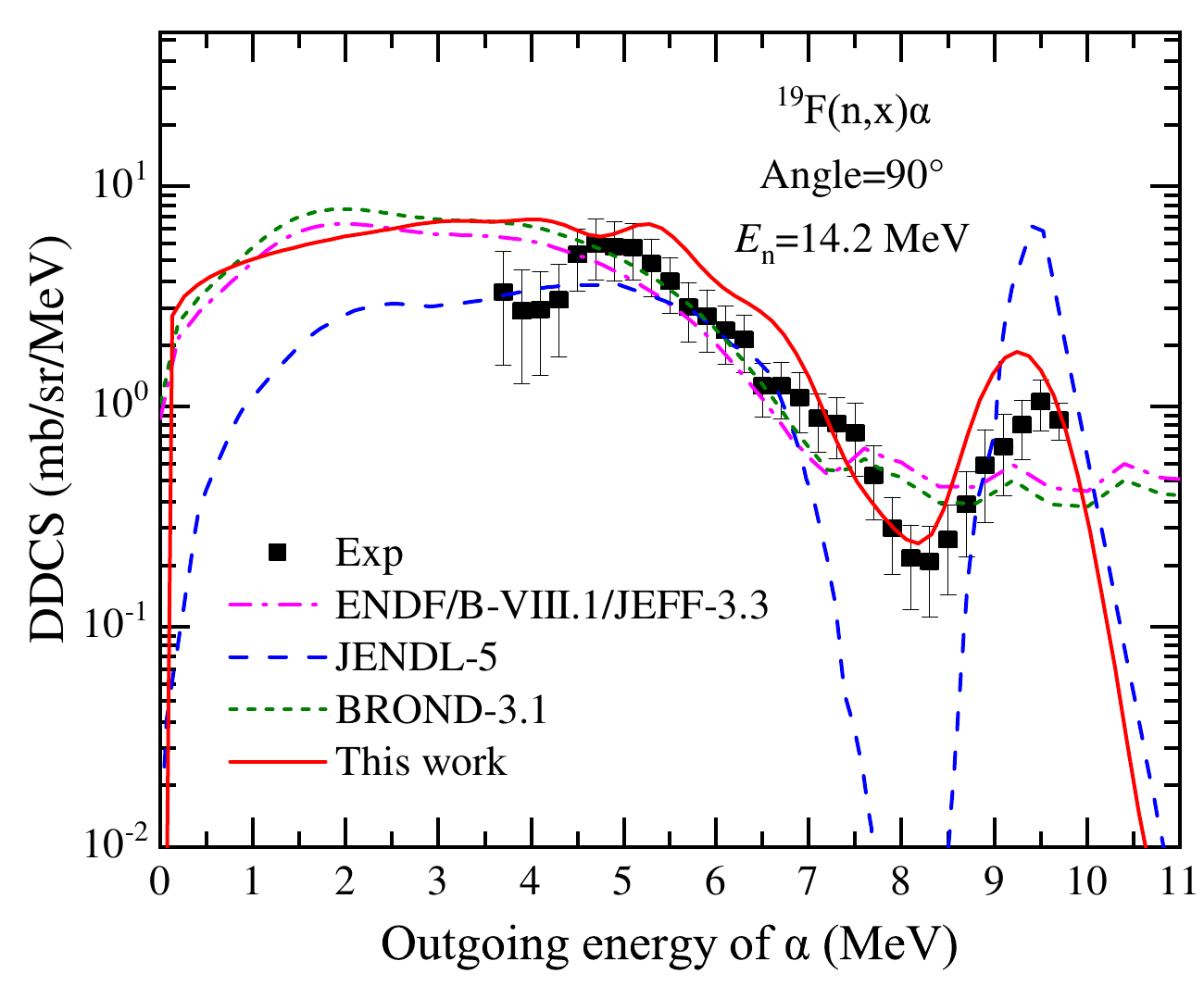}
    \caption{(Color online) Comparison of the calculated DDCS of outgoing $\alpha$ for $n + ^{19}$F reaction with the experimental data and evaluated results at $E_n$ = 14.2 MeV with $90^{\circ}$ outgoing angle. The black points denote the experimental data taken from Ref. \cite{kondo2011measurement}. The red solid, pink dash-dotted, blue dashed and green dotted lines denote the results of this work, ENDF/B-VIII.1, JENDL-5, and BROND-3.1, respectively.}
    \label{Fig16}
\end{figure}
Fig. \ref{Fig16} shows a comparison of the DDCS of outgoing $\alpha$ for $n + ^{19}$F reaction with the experimental data and evaluated results at $E_n$ = 14.2 MeV with $90^{\circ}$ outgoing angle. The black points denote the experimental data taken from Ref. \cite{kondo2011measurement}. The red solid, pink dash-dotted, blue dashed and green dotted lines denote the results of this work, ENDF/B-VIII.1, JENDL-5, and BROND-3.1, respectively. The evaluated data of ENDF/B-VIII.1, JENDL-5, and BROND-3.1 agree with the experimental DDCS in low energy region of outgoing $\alpha$. However, the results of JENDL-5 overestimate at the first peak from the right side, while the results of ENDF/B-VIII.1 and BROND-3.1 fail to account for their tendency at the same region.

\begin{figure}[!htb]
	\centering
	\includegraphics[width=0.6\linewidth]{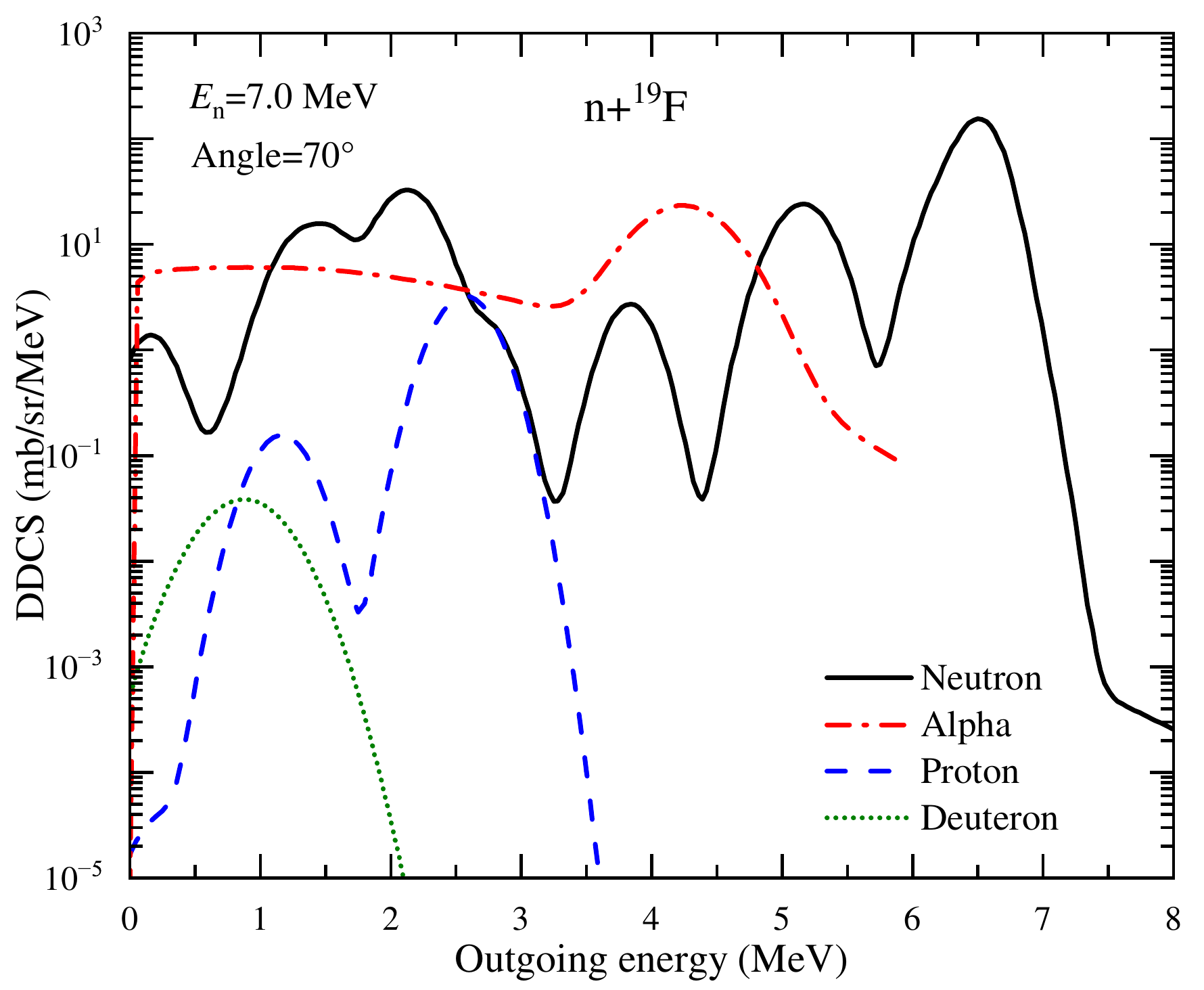}
	\caption{\textcolor{blue}{(Color online) The calculated
		total double-differential cross sections of outgoing neutron and light charged particles for $n + ^{19}$F reaction with angle of 70$^\circ$ at $E_n$ = 7 MeV in the LS. The black solid, red dash-dot, blue dash, and green dot lines denote the predicted total double-differential cross sections of neutron, alpha, proton, and deutron, respectively.}}\label{Fig17}
\end{figure}
\textcolor{blue}{The data of DDCS for $n + ^{19}$F reaction at low incident energies are of significance for the design of new-generation nuclear reactors. As an example, Fig. \ref{Fig17} shows the calculated results of the total double-differential cross sections of outgoing neutron and light charged particles for $n + ^{19}$F reaction with angle of 70$^\circ$ at $E_n$ = 7 MeV in the LS. The black solid, red dash-dot, blue dash, and green dot lines denote the predicted total double-differential cross sections of neutron, alpha, proton, and deutron, respectively. It worth mentioning that, for the $n + ^{19}$F reaction, triton emission is energetically forbidden at $E_n$ = 7 MeV. From the aforementioned figures, one can see that among all light charged particles, the double-differential cross section of $\alpha$ accounts for the largest proportion. This is consistent with the common phenomenon of the $\alpha$ cluster in light nuclei.}

\section{CONCLUSION}\label{sect4}
In our previous studies, STLN had been used to calculate the DDCS of outgoing neutron for $n + ^{19}$F reaction, and the results were good agreement with the experimental data. In this paper, the pick-up mechanism of complex particles, as one of the important components of STLN, is improved to describe the DDCS of outgoing charged particles, considering the effects of energy levels with energy, angular momentum and parity conservations. After reproducing the calculated DDCS of outgoing neutrons, the DDCS of outgoing charged particles (including $p, d, t, \alpha$) are self-consistently obtained. The results of this work are not only in good agreement with the existing measurements at $E_n$=14.2 MeV, but also superior to the data recommended by the current major nuclear databases. The results validate the effectiveness of STLN in describing the DDCS of outgoing charged particle.

\textbf{Acknowledgements}

This work was partially supported by Guangxi Key R $\&$ D Project (Guike AB24010296), Innovation Project of Guangxi Graduate Education (YCBZ2025076), and Central Government Guides Local Scientific and Technological Development Funds of China
(Guike ZY22096024).

\end{document}